\documentclass{amsart}%
\pdfoutput=1
\usepackage{amsmath}%
\usepackage{amsfonts}%
\usepackage{amssymb}
\usepackage{amsaddr}
\usepackage{graphicx}
\usepackage{natbib}
\usepackage{booktabs}
\usepackage{url}
\usepackage{listings}
\newcommand{\bse}{\boldsymbol{\eta}}
\newcommand{\bsv}{\boldsymbol{v}}

\newcommand{\bsZ}{\boldsymbol{Z}}

\newcommand{\bstheta}{\boldsymbol{\theta}}

\newcommand{\moparm}{\boldsymbol{\Psi}}

\newcommand{\bsr}{\boldsymbol{r}}

\newcommand{\ud}{\mathrm{d}}
\newcommand{\bds}[1]{\boldsymbol{#1}}
\newcommand{\te}[1]{\textrm{#1}}

\begin{document}

\author[A. T. Karl]{Andrew T. Karl}
\address{Arizona State University}
\title[The Sensitivity of College Football Rankings]{The Sensitivity of College Football Rankings to Several Modeling Choices}

\begin{abstract}
This paper proposes a multiple-membership generalized linear mixed model for ranking college football teams using only their win/loss records. The model results in an intractable, high-dimensional integral due to the random effects structure and nonlinear link function. We use recent data sets to explore the effect of the choice of integral approximation and other modeling assumptions on the rankings. Varying the modeling assumptions sometimes leads to changes in the team rankings that could affect bowl assignments.
\end{abstract}

\maketitle
\section*{Note}
This is a preprint of an article that appears in the Journal of Quantitative Analysis in Sports, Volume 8, Issue 3. The final version is available from \url{http://dx.doi.org/10.1515/1559-0410.1471}

\section{Introduction}

The highest level of collegiate football, the Football Bowl Subdivision (FBS, historically Division I-A), is undergoing a redesign of its postseason structure. Currently, the FBS is unique in that its season does not end with a tournament to decide a champion. After the regular season, the top performing teams are invited to one of several ``bowl games.'' Through 1997, the bowl assignments were strictly a function of conference membership (with special rules for non-conference teams), meaning the top two ranked teams often played in different bowl games. After the bowl games, the national champion was decided by the Associated Press and the Coaches polls. The championship was split when the polls disagreed. To prevent this, starting in 1998 the Bowl Championship Series (BCS) began to seed the Fiesta, Orange, Rose, and Sugar Bowls with certain conference champions and other highly ranked teams. The ranking procedures used by the BCS have changed since 1998, but have always been a weighted average of two polls and an average of several mathematical models, referred to as ``computer rankings''.	The computer rankings were originally allowed to use margin of victory, but in an attempt to prevent teams from running up scores to inflate their ratings, only win/loss information may now be used (ties are necessarily settled in overtime). This restriction to a binary response complicates the ranking process. Analyzing the binary game outcomes requires modeling assumptions and decisions that are not needed when modeling continuous outcomes, such as margin of victory. We explore the sensitivity of team rankings to these assumptions.

\citet{stern04} provide a detailed history of college football rankings as well as descriptions of the models currently employed by the BCS. An additional literature review is provided by \citet{west08}. \citet{stern04} discuss popular controversies surrounding the BCS system through the beginning of the 2004 season, and \citet{stern06} calls for a boycott of the BCS by quantitative analysts. Due in part to widespread criticism (and perhaps in larger part to potential television revenue), the current BCS structure will be revised to include a four team playoff beginning with the 2014 season. The tournament will be seeded by a selection committee, which may choose to factor mathematical rankings into their decision. A similar selection committee uses the Ratings Percentage Index (RPI) and proprietary models of Jeff Sagarin to help seed the NCAA Men's Division I Basketball Championship \citep{west06}.

There are several mathematical models available for ranking teams. One approach for continuous outcomes is to model a rating $\eta_i$ for each team $i$, and compute the predicted win margin of team $i$ over team $j$ via a function of the difference $\eta_i-\eta_j$. Such models require the margin of victory from past games. \citet{harville77} and \citet{harville03} develop such a model for continuous responses, and provide methods for limiting the utility of running up the score beyond a threshold $C$ by either truncating the win margins at $C$, or by scaling the margins that exceed $C$ via a hazard function. \citet{gill09} examine differences in the rankings resulting from treating the team effects as fixed or as random, as well as the modifications proposed by \citet{harville03}.

Despite the existence of models that minimize the advantage of running up the score, such as those proposed by \citet{harville03}, the BCS has decided to use only win/loss information in the computer rankings. In order to model the probability that team $i$ defeats team $j$, the Thurstone-Mosteller model \citep{thurstone,mosteller} calculates $\Phi(\eta_i-\eta_j)$, where $\Phi$ is the cumulative distribution function of the standard normal distribution. This is similar to the Bradley-Terry model \citep{bradley} used by \citet{keener}. However, all of these models encounter an infinite likelihood if any teams have a perfect record, due to quasi-complete separation of the data \citep{allison}. \citet{mease} proposes a penalized likelihood approach which circumvents the difficulty associated with the presence of undefeated or winless teams. In essence, the fixed effect model proposed by \citet{mease} becomes a random effects model with his introduction of a penalty function. Modeling the team ratings $\eta_i$ with random instead of fixed effects avoids the problem of complete or quasi-complete separation. In this case, the empirical best linear unbiased predictors (EBLUPs) of the random effects are sorted to form the team rankings.

The penalized likelihood used by \citet{mease} requires a choice of a penalty function. \citet{annis} express concern that this subjective choice may influence the rankings produced by the model.  The model proposed by \citet{mease} may be seen as a special case of the generalized linear mixed model (GLMM) we propose: it arises as one particular approximation of the marginal likelihood of our GLMM. Furthermore, \citet{mease} uses a probit link, and mentions the possibility of using a logit link as an alternative, noting that the choice between the two ``did not affect the resulting rankings substantially for the football seasons considered.''

The choice of link function is often minimized in discussions of generalized linear mixed models, and the choice of integral approximation depends on computational feasibility as determined by the structure of the random effects. It is important to note that, even if the true parameter values were known, the EBLUPs in a GLMM depend on the chosen integral approximation. Finally, it is well known that maximum likelihood (ML) estimates for variance components are subject to a downward bias, and that restricted maximum likelihood (REML) estimation procedures correct for this bias \citep{1977}.  In this paper we explore how these modeling choices affect the team rankings.

We present a GLMM for the ranking of college football teams using only win/loss information. Our model structure is the same as the one proposed by \citet{mease}, except we account for the teams via random instead of fixed effects. Furthermore, we consider treating the FBS and FCS (Football Championship Subdivision, formerly Division I-AA) divisions as different populations, whereas \citet{mease} consolidates all FCS teams into a single effect that is regarded as member of the FBS population. We show that our GLMM is a generalization of the penalized likelihood proposed by \citet{mease}, and explore the sensitivity of the rankings from our model to the choices of
\begin{enumerate}
\item link function
\item integral approximation used for the marginal likelihood of our GLMM
\item distribution of the random team effects
\item consolidating FCS teams into a single ``team'' vs. treating FCS as a separate population
\item ML vs. REML
\end{enumerate}

The results from our analysis show that the changes in team ratings resulting from varying these assumptions are small relative to the standard errors associated with each team's rating. However, these changes result in a reordering of the team rankings that may lead to practically significant differences in the bowl selection process. There is limited information in the binary win/loss outcome of each game, and it is not surprising that changing properties of the model may lead to teams with similar records swapping ranks. For some of the assumptions listed above, especially points 1 and 3, there may not be a clear set of best choices. However, these choices may affect the ordering of the team rankings. 

We present the model in Section \ref{sec:model} and discuss the parameter estimation in Section \ref{sec:estimation}. In Section \ref{sec:application}, we compare the rankings resulting from our model under different assumptions for the seasons 2008-2011. In each year, we use data through the end of the conference championships, excluding the outcomes of the bowl games. Thus our examples use the same data that were used to compile the final BCS rankings in each year, which were used as a basis for the bowl invitations.

\section{The Model}\label{sec:model}

\cite{mease} models team ratings with fixed effects $\bstheta=(\theta_1,\ldots,\theta_{p+1})$ via the likelihood
\begin{align}
l(\bstheta)=&\prod_{(i,j)\in S}[\Phi(\theta_i-\theta_j)]^{n_{ij}}\label{m1}\\
&\times\prod_{i=1}^p\Phi(\theta_i)\Phi(-\theta_i)\label{m2}\\
&\times \Phi(\theta_{p+1})\Phi(-\theta_{p+1}) \times \prod_{(i,j)\in S^*}[\Phi(\theta_i-\theta_j)]^{n_{ij}}\label{m3}
\end{align}
where $S$ is the set of all ordered pairs $(i,j)$ for which team $i$ defeated team $j$, and both teams belong to the FBS. $p$ is the number of FBS teams, and $\theta_{p+1}$ is a single effect that is used to represent every FCS team. $S^*$ is defined in the same way as $S$, except one of the teams in each pair is from the FBS and the other is from the FCS. \citet{mease} refers to Equations (\ref{m1})-(\ref{m3}) as Parts 1, 2, and 3, respectively. Part 1 models the probability of the outcome of each game using the team ratings, and implicitly considers ``strength of schedule.'' The second part is a penalty function that allows the model to be estimated in the presence of undefeated or winless teams: using Part 1 alone leads to an infinite likelihood in these cases. The third and final part models games between FBS and FCS teams using a single team effect to represent all FCS teams.

We propose modeling the team ratings with random instead of fixed effects, and show that the model proposed by \citet{mease} is actually a special case of our random effects model. Treating the teams as random effects requires the specification of a distribution for those effects, as well as the choice of an integral approximation for the resulting generalized linear mixed model. In addition, we propose another method of modeling games between FBS and FCS teams. Besides modeling all FCS teams as a single effect and ignoring FCS games that were not played against an FBS opponent, we model FCS teams as a separate population. Using this approach, we include all of the games played between two FCS teams, but ignore FCS games played against lower-divisional opponents. This introduces two more modeling assumptions. 1) Instead of ignoring FCS games against lower divisional opponents, those opponents could be treated as a single effect, using the same approach as \citet{mease}. Although we do not take this approach, this could protect against the possibility of a successful FCS team being overrated due to an ignored loss against a Division II team. 2) When modeling separate FBS and FCS populations, we may either assume that the populations share a common variance, or we may model a different variance for each population. In Section \ref{sec:application}, we model both pooled and separate variances and compare the resulting differences.

\subsection{Separate FCS Population with Pooled Variance}\label{ssec:fe.p.1}

We first present our model including separate FBS and FCS populations, assuming that the FBS and FCS distributions share a common variance. In Section \ref{ssec:fe.p.2} we describe the changes necessary for modeling different FBS and FCS effect variances, and in Section \ref{ssec:fe.p.0} we treat FCS opponents of FBS teams as a single team in the FBS population

Our model considers outcomes of FBS and FCS games in a given season. Let $r_i$ be a binary indicator for the outcome of the $i$-th game for $i=1,\ldots,n$, taking the value 1 with a home team win and 0 with a visiting team win. For neutral site games, designate a home team arbitrarily. 

We model the rating of the $j$-th team for $j=1,\ldots,p+q$ with a random effect $\eta_j$ assuming $\eta_j\sim N(0,\sigma^2_t)$. $p$ and $q$ represent the number of included FBS and FCS teams, respectively. We assume that the distributions of FBS and FCS ratings share a common variance $\sigma^2_t$, but that the distributions have different means. We account for the difference in means between the two divisions by including the fixed effect $\beta$ in the model. The coefficient $X_i$ for $\beta$ takes the value 1 if the $i$-th game involves an FCS team visiting an FBS team, and 0 otherwise (FBS teams do not travel to play FCS teams). We will refer to $\beta$ as the ``FCS effect.''

Using the threshold model of \citet{mcculloch94}, we assume that the game outcomes are determined by a latent continuous variable $y_i=X_i\beta+\bsZ_i\bse+\epsilon_i$, which may be interpreted as the margin of victory for the home team, but only the binary outcome $r_i=I_{\{y_i>0\}}$ is observed.  We thus model the probability $\pi_i$ of a home team win in the $i$-th game, as.
\[\pi_i=P(X_i\beta+\bsZ_i\bse+\epsilon_i>0) \]
where $\beta$ is the FCS effect, $\bse=(\eta_1,\ldots,\eta_{p+q})\sim N(0,\sigma^2_t\bds{I})$ contains the random effects representing the team ratings, $\bds{\epsilon}=(\epsilon_1,\ldots,\epsilon_n)\sim N(0,\bds{I})$, and $cov(\bds{\eta},\bds{\epsilon})=0$. The assumed distribution of $\epsilon$ determines the link function of the resulting GLMM. Assuming that $\bds{\epsilon} \sim N(0,\bds{I})$ leads to a probit link, 
\begin{align*}
r_i|\bse&\sim \te{Bin}(1,\pi_i)\\
\Phi^{-1}(\pi_i)&=X_i\beta+\bsZ_i\bse
\end{align*}
By contrast, assuming that the $\epsilon_i$ are independent and follow a logistic distribution leads to a logit link.
\begin{align*}
r_i|\bse&\sim \te{Bin}(1,\pi_i)\\
\te{logit}(\pi_i)&=X_i\beta+\bsZ_i\bse
\end{align*}
where $\te{logit}(\pi)=\log(\pi/(1-\pi))$.

The design matrix $\bsZ$ for the random effects contains rows $\bsZ_i$ that indicate which teams competed in game $i$. If team $k$ visits team $l$ in game $i$, then $\bsZ_i$ is a vector of zeros with a $1$ in position $l$ and a $-1$ in position $k$. This is a multi-membership design \citep{browne01} since each game belongs to multiple levels of the same random effect. As a result, $\bsZ$ does not have a patterned structure and may not be factored as it could be with nested designs. \citet{mease} uses the same design matrix, albeit using fixed effects for the teams. The multi-membership $\bsZ$ matrix is implied by the structure of the product over the set $S$ in Equation (\ref{m1}).

We choose not to include a home field effect. We are concerned about the potential impact of nonrandom scheduling between teams on the estimate of home field advantage. Only three of the 120 FBS college football teams (Notre Dame, UCLA, and Southern California) have refrained from scheduling matches against FCS teams to lighten their schedules. These games are accounted for with the FCS effect, but there is yet another concern about the estimation of a home field effect. The more competitive programs tend to play more home games than other FBS programs. These additional home games are often against low-level FBS teams who do not posses the bargaining leverage to request a home-and-home series.  As a result, the ``home field advantage'' may appear to be significant simply because of the tendency of larger schools to schedule a number of easier home games. Others have advocated including a home field effect, including \citet{harville77}, \citet{harville03}, \citet{mease}, and \citet{wang}. 

Other fixed effects may easily be included in the model, but the factors should not contain any levels that act as perfect predictors for the outcomes. For example, the inclusion of the FCS fixed effect will lead to an infinite likelihood due to quasi-complete separation \citep{allison} in the event that every FBS vs. FCS game results in an FBS win. In 2008, FCS teams won only 2 of 86 games against FBS teams. In the event that FBS teams were to win all such games in a year, the FCS effect and teams may be removed from the model. Historically, some of the BCS computer rankings always ignored these inter-division matches, until fifth-ranked Michigan lost to FCS Appalachian State after paying the team \$400,000 for a one-off home game in 2007.

The likelihood for our model with a probit link is
\begin{equation}\label{eq:probit}
L(\beta,\sigma^2_t)=\idotsint \prod_{i=1}^n \left[\Phi\left(\left(-1\right)^{1-r_i}\left[X_i\beta+\bsZ_i\bse\right]\right)\right] f(\bse) \ud \bse
\end{equation}
where $f(\bse)$ is the density of $\bse$. We assume $\bse\sim N_{p+q}(0,\sigma^2_t \bds{I})$. Using a logit link,
\begin{equation}\label{eq:logit}
L(\beta,\sigma^2_t)=\idotsint \prod_{i=1}^n \left(\frac{e^{X_i\beta+\bsZ_i\bse}}{1+e^{X_i\beta+\bsZ_i\bse}}\right)^{r_i} \left(\frac{1}{1+e^{X_i\beta+\bsZ_i\bse}}\right)^{1-r_i}f(\bse) \ud \bse.
\end{equation}

The model likelihood functions in Equations (\ref{eq:probit}) and (\ref{eq:logit}) contain intractable integrals because the random effects enter the model through a nonlinear link function. Furthermore, the ($p+q$)-dimensional integral in each equation may not be factored as a product of one-dimensional integrals. Such a factorization occurs in longitudinal models involving nested random effects. However, the multi-membership random effects structure of our model results in a likelihood that may not be factored.

\subsection{Separate FCS Population with Unique Variance}\label{ssec:fe.p.2}
The model assuming separate FBS and FCS populations uses the same setup as described in Section \ref{ssec:fe.p.1}, except that the populations are allowed to have different variances. The likelihood function for this model with a probit link is
\begin{equation*}
L(\beta,\sigma^2_1,\sigma^2_2)=\idotsint \prod_{i=1}^n \left[\Phi\left(\left(-1\right)^{1-r_i}\left[X_i\beta+\bsZ_i\bse\right]\right)\right] f(\bse) \ud \bse
\end{equation*}
where $f(\bse)$ is the density of $\bse=(\bse_1,\bse_2)$, with $\bse_1$ and $\bse_2$ containing the FBS and FCS team effects, respectively. We assume $\bse_1\sim N_p(0,\sigma^2_1 \bds{I})$, $\bse_2\sim N_q(0,\sigma^2_2 \bds{I})$, and that $cov(\bse_1,\bse_2)=0$.
Using a logit link,
\begin{equation*}
L(\beta,\sigma^2_1,\sigma^2_2)=\idotsint \prod_{i=1}^n \left(\frac{e^{X_i\beta+\bsZ_i\bse}}{1+e^{X_i\beta+\bsZ_i\bse}}\right)^{r_i} \left(\frac{1}{1+e^{X_i\beta+\bsZ_i\bse}}\right)^{1-r_i}f(\bse) \ud \bse.
\end{equation*}

\subsection{Single Population}\label{ssec:fe.p.0}
The model consolidating FCS teams into a single ``team'' in the FBS population also uses a similar setup as described in Section \ref{ssec:fe.p.1}. However, this model does not require the FCS fixed effect, and discards information about games played between pairs of FCS teams. The likelihood function using a probit link is
\begin{equation}\label{eq:ourmodel}
L(\sigma^2_t)=\idotsint \prod_{i=1}^n \left[\Phi\left(\left(-1\right)^{1-r_i}\bsZ_i\bse\right)\right] f(\bse) \ud \bse
\end{equation}
where $f(\bse)$ is the density of $\bse$. We assume $\bse\sim N_{p+1}(0,\sigma^2_t \bds{I})$, where $\eta_{p+1}$ is consolidated FCS team-effect.
Using a logit link,
\begin{equation*}
L(\sigma^2_t)=\idotsint \prod_{i=1}^n \left(\frac{e^{\bsZ_i\bse}}{1+e^{\bsZ_i\bse}}\right)^{r_i} \left(\frac{1}{1+e^{\bsZ_i\bse}}\right)^{1-r_i}f(\bse) \ud \bse.
\end{equation*}

The model proposed by \citet{mease} results from applying a particular integral approximation to Model (\ref{eq:ourmodel}). Following an illustration by \citet{demidenko}, the penalized likelihood used by \citet{mease} may be derived from our model (\ref{eq:ourmodel}) via the Laplace approximation \citep{evans95}. Letting
\[h(\bse)=\log\left[\prod_{i=1}^n \left[\Phi\left(\left(-1\right)^{1-r_i}\bsZ_i\bse\right)\right] f(\bse)\right],\]
the Laplace approximation yields
\begin{equation} \label{eq:approx}
L(\sigma^2_t)\approx (2\pi)^{n/2}e^{h(\bse^*)}\left|\left. -\frac{\partial^2 h}{\partial \bse\partial \bse^{\prime}}\right|_{\bse=\bse^*}\right|^{-1/2}
\end{equation}
where $\bse^*$ is the mode of $h(\bse)$. Further assuming that the determinant in Equation (\ref{eq:approx}) varies slowly with $\bse$ yields the quasi-likelihood \citep{breslow93}
\begin{equation}\label{eq:approx2}
L(\bse)\approx \prod_{i=1}^n \left[\Phi\left(\left(-1\right)^{1-r_i}\bsZ_i\bse\right)\right] f(\bse)
\end{equation}
If the random effects $\bse$ are assumed to be distributed so that
\begin{equation}\label{eq:penalty}
f(\bse)\propto\prod_{j=1}^{p+1}\Phi(\eta_j)\Phi(-\eta_j),
\end{equation}
 then Equation (\ref{eq:approx2}) yields the likelihood presented by \citet{mease} in Equations (\ref{m1}-\ref{m3}). Thus the model of \citet{mease} may be viewed as the PQL approximation to our probit model (\ref{eq:ourmodel}), where the random effects are assumed to have the density specified in Equation (\ref{eq:penalty}) rather than the normal density that we specified.

\section{Parameter Estimation}\label{sec:estimation}
Section \ref{ssec:fe.p.0} demonstrates that the model likelihood of \citet{mease} is the PQL approximation to our model (\ref{eq:ourmodel}) under a certain non-normal distribution of the random effects, $\bse$. PQL is based on the Laplace approximation, but makes one further approximation, and produces biased parameter estimates \citep{breslow95}. By contrast, the Laplace approximation produces consistent estimators since the dimension of our integrals is equal to the number of teams and does not increase with the sample size \citep{shun95}. In Section \ref{sec:estimationsas} we demonstrate how our model may be fit in SAS using a PQL approximation, and in Section \ref{sec:estimationem} we explain how we use an EM algorithm with both a first-order and a fully exponential Laplace approximation to obtain the rankings. Code for fitting the model proposed by \citet{mease} is available from \citet{measeweb}.

\subsection{Fitting the Model in SAS}\label{sec:estimationsas}
Specifying the multi-membership random effects structure in SAS is possible in PROC GLIMMIX via the MULTIMEMBER option of the EFFECT statement. However, GLIMMIX does not take into account the fact that $\bsZ$ is sparse. This is not an issue for the football data sets used here, but becomes problematic in other settings with larger data sets. The EM algorithm we use in Section \ref{sec:estimationem} provides computational advantages in the estimation of multi-membership models 
\citep{karlem}.

The default estimation method of PROC GLIMMIX is a doubly-iterative pseudo-likelihood method by which the link function is linearized and a linear mixed model is fit at each iteration. This method is equivalent to PQL when the scale parameter is set to 1, which is the default setting for the Bernoulli distribution in PROC GLIMMIX \citep{wolfinger93, sasbook}. Example SAS code appears in Appendix \ref{sec:sascode}.

\subsection{Fitting the Model with an EM Algorithm}
\label{sec:estimationem}
The EM algorithm \citep{dempster77,embook} is often used for maximum likelihood estimation in the presence of missing data. It may be used for the estimation of mixed models by treating the random effects as missing observations \citep{laird82}, although an integral approximation is necessary when the random effects enter the model through a nonlinear link function, such as is the case with our model. Note that the high dimension of the integral renders quadrature methods infeasible.

The use of a fully exponential Laplace approximation \citep{tierney89} with an EM algorithm for the estimation of generalized linear mixed models was first proposed by \citet{steele}. \citet{riz09} use this method to estimate the parameters of a joint model for a continuous longitudinal process and a time-to-dropout measurement. \citet{karlphd} applies this approach to a multi-membership joint model. We will give a brief overview of the EM estimation procedure, and refer to \citet{karlphd} for further details, as well as to \citet{riz09} for similar calculations made in the setting of nested random effects.

As an alternative to the PQL approximation, we fit the model using an EM algorithm with custom-written code in R \citep{R} with both a first order Laplace (LA) and a fully exponential Laplace (FE) approximation in the E-step. The Laplace approximations are more accurate than PQL, but are more computationally demanding, requiring the calculation of higher-order derivatives than PQL. The first order Laplace approximation requires the first two derivatives of the integrand. Calculation of the fully exponential Laplace approximation for the conditional mean of the random effects requires the third derivative, and calculation of the conditional variance requires the fourth derivative. The approximation is complicated by the multi-membership random effects structure. 

We outline the EM procedure for only one of the models, and note that the calculations are similar for the other models. To estimate the parameters $\moparm=(\beta,\sigma_t^2)$ of the model in Equation (\ref{eq:probit}), we use the equations derived by \citet{karlphd}, ignoring the longitudinal process in the joint model he analyzed. Given initial values for the parameters and the random effects, the EM algorithm alternates between an expectation (E) step and a maximization (M) step. At iteration (k + 1), the E step calculates the conditional expectation of the log-likelihood $\log f(\bsr, \bse)$,
\begin{equation*}
Q(\moparm; \moparm^{(k)}) = \idotsint {\left\{\log f\left(\bsr|\bse; \moparm\right) + \log f\left(\bse; \moparm\right)\right\} f(\bse|\bsr; \moparm^{(k)})} \ud\bse,
\end{equation*}
given the vector of game outcomes, $\bsr$, and parameter estimates $\moparm^{(k)}$ obtained in the k-th iteration. For this calculation, it is sufficient to find the conditional mean $\widetilde{\bse}=\te{E}[\bse|\bsr;\moparm^{(k)}]$ and the conditional variance $\widetilde{\bsv}=\te{var}[\bse|\bsr;\moparm^{(k)}]$ of the random effects. The M-step then maximizes $Q(\moparm; \moparm^{(k)})$
with respect to $\moparm$, resulting in the updated parameter vector $\moparm^{(k + 1)}$. 

The expressions for $\widetilde{\bse}$ and $\widetilde{\bsv}$ involve intractable integrals, necessitating the use of the Laplace approximations. The gradient and inverse Hessian of the joint distribution $f(\bsr,\bse)$ (with respect to $\bse$) furnish the first order Laplace approximations to $\widetilde{\bse}$ and $\widetilde{\bsv}$, respectively, and the fully exponential Laplace approximations require computationally expensive corrections to these values. Upon convergence, $\widetilde{\bse}$ serves as the vector of team ratings, demonstrating that the choice of integral approximation affects the ratings even if the model parameters $\moparm$ are known.

The M-step update for $\widehat{\beta}$ from the probit model may be obtained by setting the score function
\begin{align}
S(\widehat{\beta})
=&\sum_{i=1}^n X_i\idotsint{\left(-1\right)^{1-r_{i}}\frac{\phi\left[\left(-1\right)^{1-r_{i}}\left(X_i\widehat{\beta}+\bsZ_{i}
\bse\right)\right]}{\Phi\left[\left(-1\right)^{1-r_{i}}\left(X_{i}\widehat{\beta}+\bsZ_{i}
\bse\right)\right]}}f(\bse|\bsr)\ud \bse \label{eq:mbeta}
\end{align}
equal to 0, where $\phi$ is the density function of the standard normal distribution. The equation may be solved via Newton-Raphson, using a central difference approximation to $S(\widehat{\beta})$ in order to obtain the necessary Hessian. Fortunately, there is a closed form M-step update for $\sigma^2_t$, namely
\[\widehat{\sigma}^2_t=\frac{1}{p+q}\te{trace}(\widetilde{\bsv}+\widetilde{\bse}\widetilde{\bse}^{\prime}). \]

\section{Application}\label{sec:application}
We obtain the game outcomes for the 2008-2011 seasons from the NCAA website \citep{ncaaweb}. The data require additional processing, since outcomes are recorded by teams, resulting in duplicate observations for games between two teams within the same division. Some of the neutral site games were duplicated while others were not. We combine the FBS and FCS files, remove all games labeled ``away'', remove games between FCS and lower division teams, purge redundant neutral site games, add an indicator for games played between FBS and FCS schools, and remove the records of games played after the production of the final BCS rankings in each year. The processed data are available from \citet{karlweb}. See Table \ref{tab:data} in Appendix \ref{sec:sascode} for the first observations of our 2008 data set. 

The team ratings and rankings for the 2008-2011 seasons appear in tables and scatter plots Appendix \ref{sec:tables}. The scatter plots are printed with reference lines with slope 1 and intercept 0. The rankings use the game outcomes through the end of the conference championships in each year: the bowl outcomes are not included. This allows us to compare the rankings from our model to the final BCS rankings used to choose the BCS teams. The BCS rankings are included as a reference, and not as a standard that we expect our model to match. Two-thirds of the weight of the BCS ranking is given to polls, and the voters are allowed to consider more than each team's win/loss record.

Under the current BCS configuration, the top 16 teams of the BCS rankings are relevant due to the rules involving eligibility for selection in the non-championship BCS bowls. We do not list all of the rules the selection process, but instead point out a few of the highlights. The top two teams in the BCS rankings are selected to play in the national championship game. The remaining 8 BCS slots are filled with the winners of certain conferences (e.g. Big East), regardless of their ranking. For example, after the 2010 season, \#7 ranked Oklahoma was paired in the Fiesta Bowl with an (8-4) unranked Connecticut team. 

There are special rules for non-BCS teams, including a rule that the highest-ranked winner of a non-BCS conference will receive a berth if it is either ranked in the top 12 or in the top 16 and higher than at least one BCS conference winner. Under certain conditions, any team finishing in the top four is guaranteed a berth. The complete list of rules is available from the \citet{bcs}. In short, permutations in the rankings of teams outside of the top two can affect the selection of teams for participation in the BCS. The value of these berths is substantial. Each conference, and thus each school, receives an extra payout for each additional team that is awarded a BCS berth. The head coaches of these teams benefit as well, due to their contracts: Les Miles (LSU) received a \$200,000 bonus for reaching a BCS game in 2012 and would have received an extra \$100,000 for winning the national title game. In addition, a national title win for Miles would have activated a clause in his contract giving him a \$5.7 million raise over the remaining 6 years of his contract. Instead, Nick Saban (Alabama) received a \$400,000 bonus for defeating LSU \citep{bloomberg}.

In the following subsections, we describe the changes in rankings that result from varying several modeling assumptions. For convenience, we will introduce a three-part notation to indicate the method of integral approximation (PQL, LA for Laplace, FE for fully exponential Laplace), the link function (P for probit, L for logit), and the way in which FCS teams are handled (0 for consolidating them into a single effect as in Section \ref{ssec:fe.p.0}, 1 for modeling separate FBS and FCS populations with a pooled variance as in Section \ref{ssec:fe.p.1}, and 2 for modeling separate populations with separate variances as in Section \ref{ssec:fe.p.2}). For example, PQL.P.0 denotes the PQL approximation to the probit model that consolidates FCS teams into a single ``team.''

\subsection{The Link Function}\label{ssec:link}
To explore sensitivity to the choice of link function, we compare the rankings from FE.P.0 and FE.L.0, which appear in Figures \ref{plot:2008_d}, \ref{plot:2008_c}, \ref{plot:2009_d}, \ref{plot:2009_c}, \ref{plot:2010_d}, \ref{plot:2010_c}, \ref{plot:2011_d}, and  \ref{plot:2011_c} in Appendix \ref{sec:tables}. The reference lines in the scatter plots for the continuous ratings from these two models are not meaningful since the kurtosis of the logistic distribution is greater than that of the normal distribution. We would expect the rankings using the fully exponential Laplace approximation to be the most sensitive to the choice of link function, since the approximation makes use of four derivatives of the complete-data likelihood, and thus the link function. 

\begin{figure}
\caption{FE.P.0 Ratings with Standard Errors for 2008 Season.}
\label{plot:2008_ratings}
\centering
\includegraphics[scale=.4]{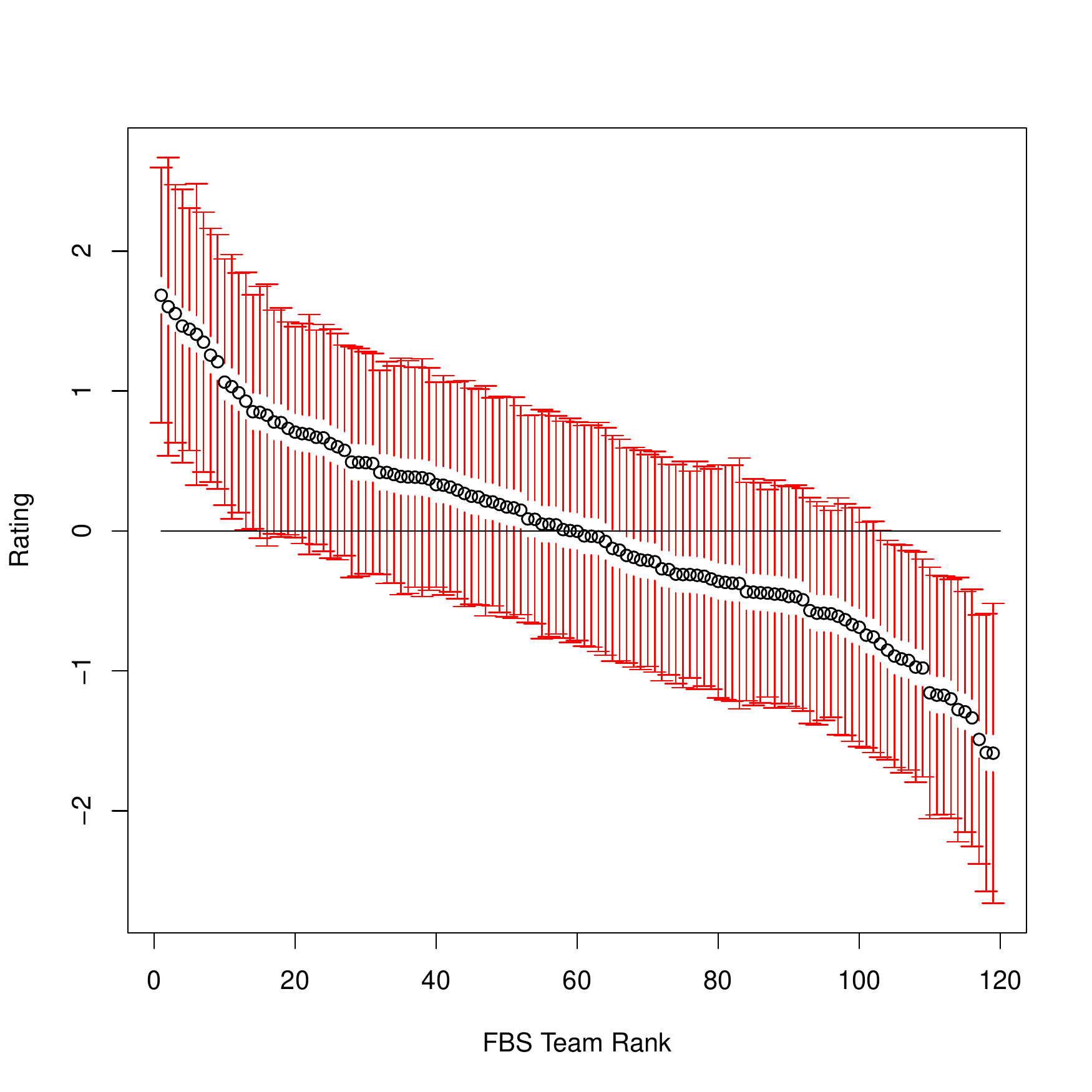}
\end{figure}

In 2008 these two models agree on ranks 1-16, though 17-20 are scrambled by one or two positions. In 2009, 13-14 and 17-18 each swap positions. In 2010, 7 and 9 swap, and 17-20 each differ by one position. However, in 2011, FE.P.0 ranks LSU, Alabama, and Oklahoma St.~as the top three teams, respectively, while FE.L.0 picks LSU, Oklahoma St., Alabama. Figure \ref{plot:2011_c} shows that Alabama and Oklahoma St.~have nearly identical ratings in each of the models. The change in link function leads to slight changes in the team ratings, prompting shifts in the rankings. These changes are small relative to the standard errors of the ratings, as can be seen from the caterpillar plot in Figure \ref{plot:2008_ratings}, which displays the 2008 FBS team ratings from FE.P.0 along with their associated $95\%$ prediction intervals. 

In the context of teacher evaluation, \citet{draper95} urges caution when using the EBLUP
rankings because individual ratings may have large standard errors. Given the high stakes involved in the BCS rankings and the way the models are used by the BCS, small changes in the ratings may be practically significant, even if they are not statistically significant. The BCS uses the discretized rankings from each model, not the continuous ratings. The plots in Appendix \ref{sec:tables} show how uneven the spacing between team ratings can be. Teams with similar ratings are more likely to swap positions with changes in the modeling assumptions. 

\subsection{The Integral Approximation}\label{ssec:integral}
The tables in Appendix \ref{sec:tables} contain the team rankings and ratings, as well as the parameter estimates for $\sigma^2_t$ from models PQL.P.0, LA.P.0, and FE.P.0. The downward bias in the PQL estimates of $\sigma^2_t$ described by \citet{breslow95} is clearly present. The fully exponential Laplace approximation is more accurate than the first order Laplace, which is in turn better than PQL. It is interesting to see in the tables that the changes in a team's ranking from PQL to LA to FE are monotonic. That is, for 2008-2011, if a team is ranked X by PQL.P.0 and Y by FE.P.0, its ranking under LA.P.1 is somewhere between X and Y, inclusive.

Two of the more interesting changes in ranking seen across the models for these data are in 2009 and 2011, where the change in integral approximation alters which team is voted second, and thus one of the two teams to play for the national championship. In 2009, FE.P.0 lists Alabama and Cincinnati, while PQL.P.0 and LA.P.0 select Alabama and Texas. In 2011, PQL.P.0 and LA.P.0 rank Oklahoma St.~2, while FE.P.0 ranks Alabama 2. Of course, these teams only moved one position between second and third in these rankings, but the exchange is remarkable in the sense that it arises solely from using a fully exponential (asymptotically analogous to a second-order) instead of a first-order Laplace approximation. At least in these years, our answer to the question, ``Who do you think should play in the BCS championship game?''~depends on the response to the question, ``To what order would you prefer to extend your Laplace approximation?''

Changing from PQL.P.0 to FE.P.0 in 2008 moves Texas Tech from 6 to 4, Boise St.~from 5 to 6, Oklahoma St from 16 to 14, plus seven other changes of one position. Besides the Alabama/Cincinnati swap in 2009, four other teams change rank: Oregon St.~moves from 23 to 20, Arizona from 21 to 19, and LSU switched 12 for 11 with Virginia Tech. In 2010, Stanford improves from 5 to 4, displacing Oklahoma. Wisconsin moves from 8 to 6, Ohio St.~from 6 to 7, Arkansas from 10 to 8, Michigan St.~from 7 to 9, and Boise St.~from 9 to 10. Finally, in addition to the Alabama/Oklahoma St.~switch, the change to the fully exponential model in 2011 lifts Arkansas from 8 to 6. Baylor moves from  16 to 14, Oklahoma from 14 to 12, Michigan from 13 to 15, Georgia from 17 to 16, and Wisconsin from 15 to 17. There were a couple of other teams that changed a single rank.

It is difficult to tell whether there is a pattern in set of teams which are affected by the choice of FE.P.0 over PQL.P.0 in each year. SEC teams tend to benefit from the change. It is possible that the increased random team-effect variance estimated by the fully exponential model implicitly places a greater emphasis on strength of schedule, since the larger the variance of the team effect, the greater the estimated disparity between FBS teams. We consider this point further in Section \ref{ssec:dist}.

\subsection{Distribution(s) of the Random Team-Effects}\label{ssec:dist}

\begin{figure}
\caption{The random effects distribution (dashed) implicitly assumed by \citet{mease} and the $N(0,0.57)$ distribution from PQL.P.0 in 2009 (solid). }
\label{plot:dist_comp}
\centering
\includegraphics[scale=.4]{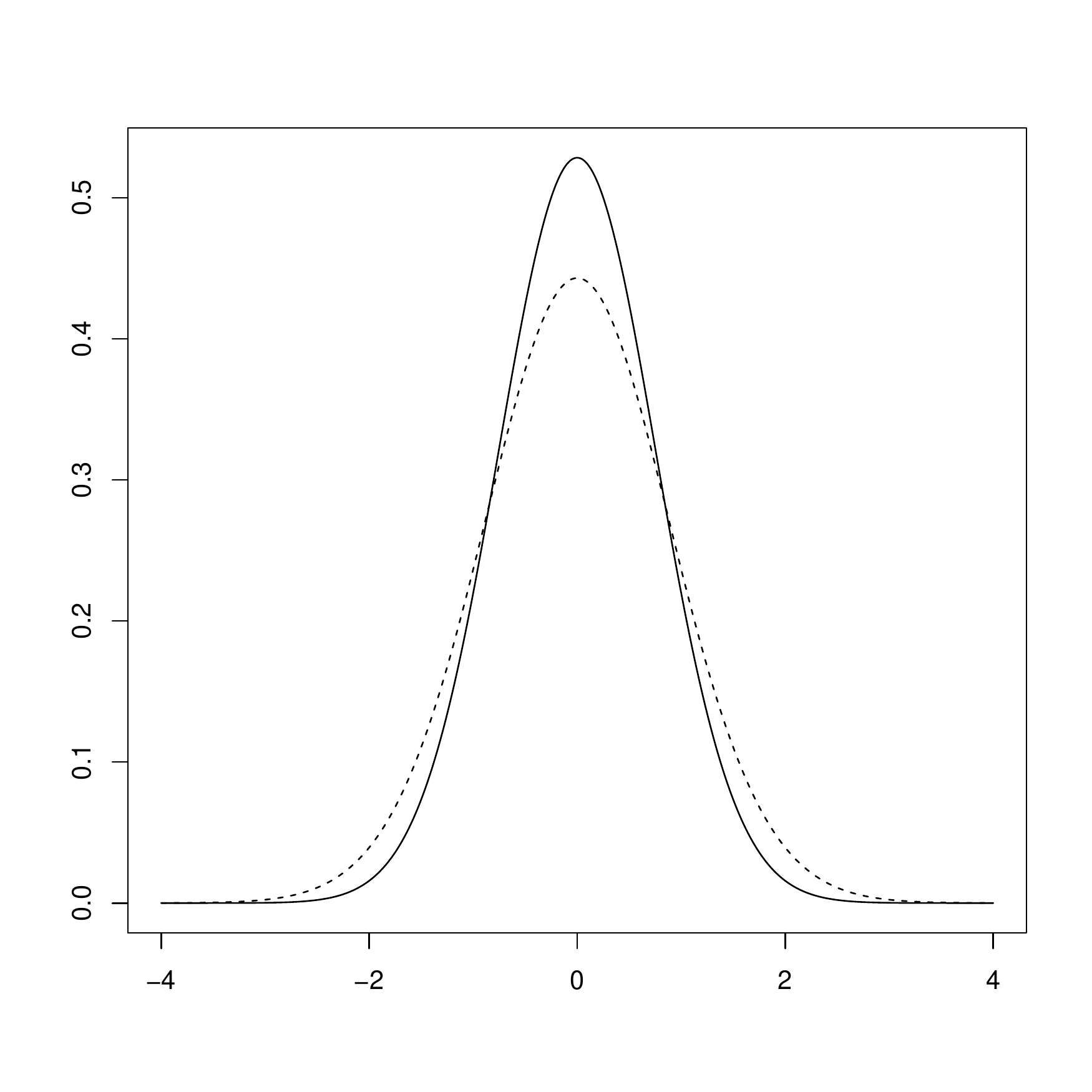}
\end{figure}

\begin{figure}
\caption{Comparison of rankings resulting from normally distributed random effects (PQL.P.0) to those obtained under the distribution assumed by \citet{mease}}
\label{plot:dist_comp2}
\centering
\includegraphics[scale=.35]{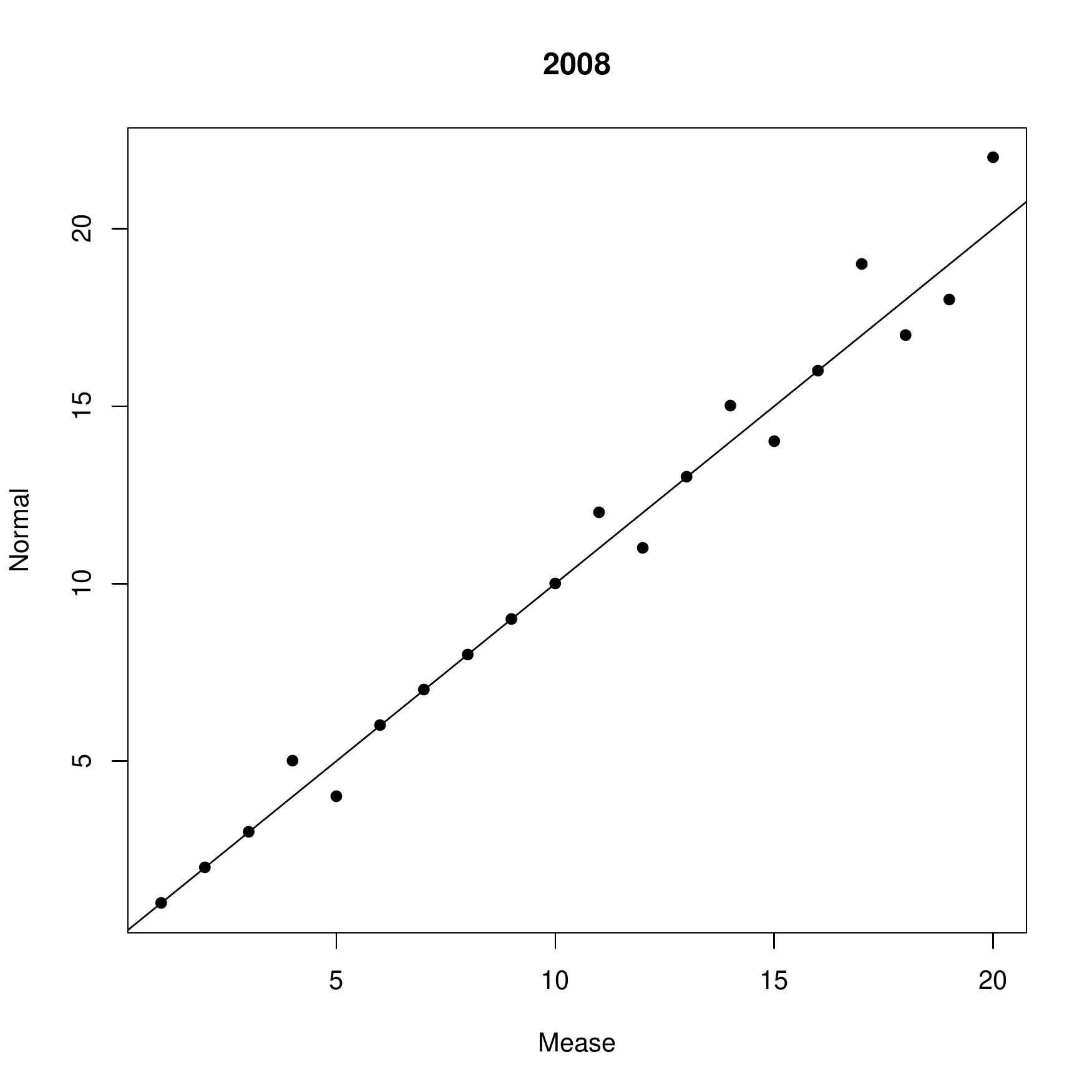}
\includegraphics[scale=.35]{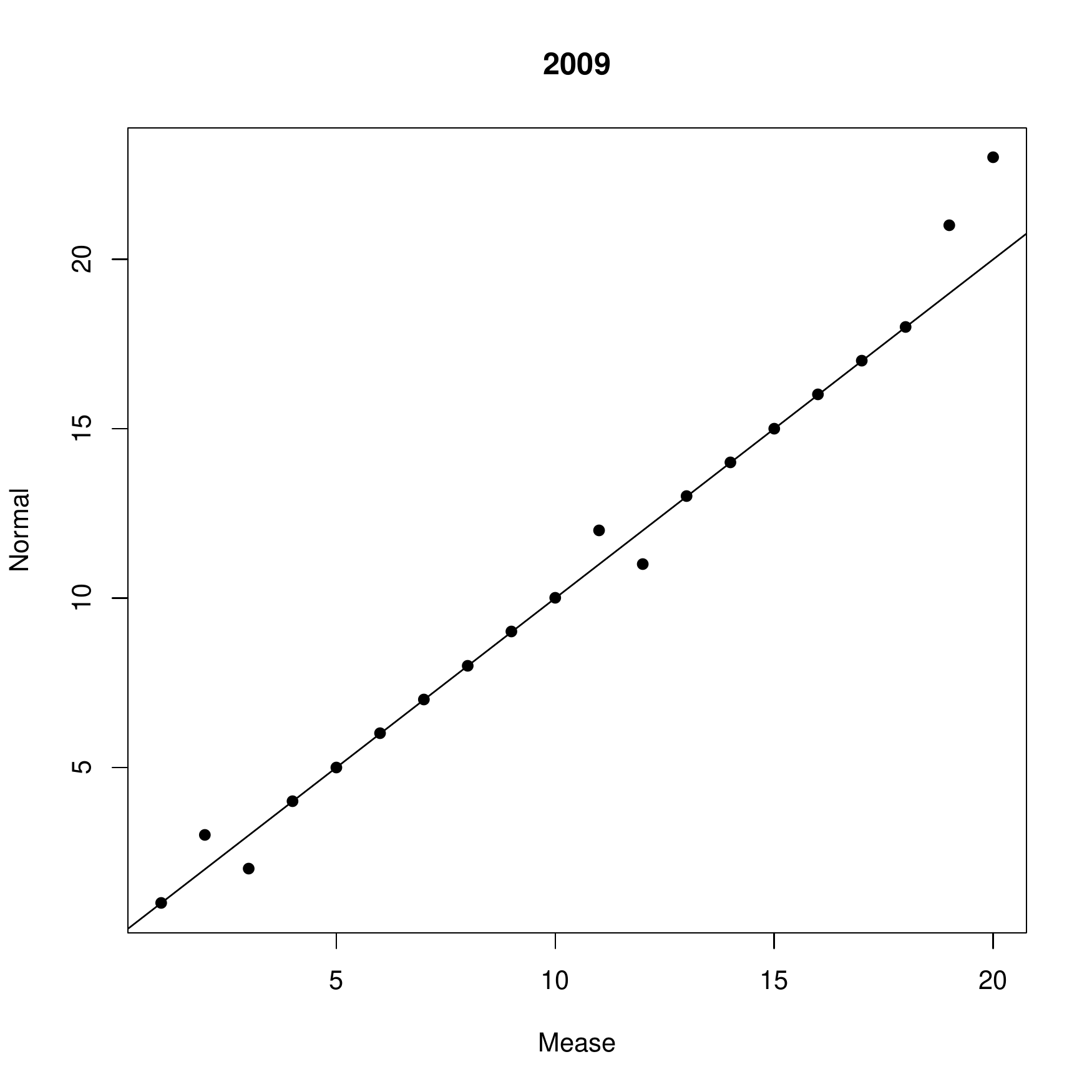}
\includegraphics[scale=.35]{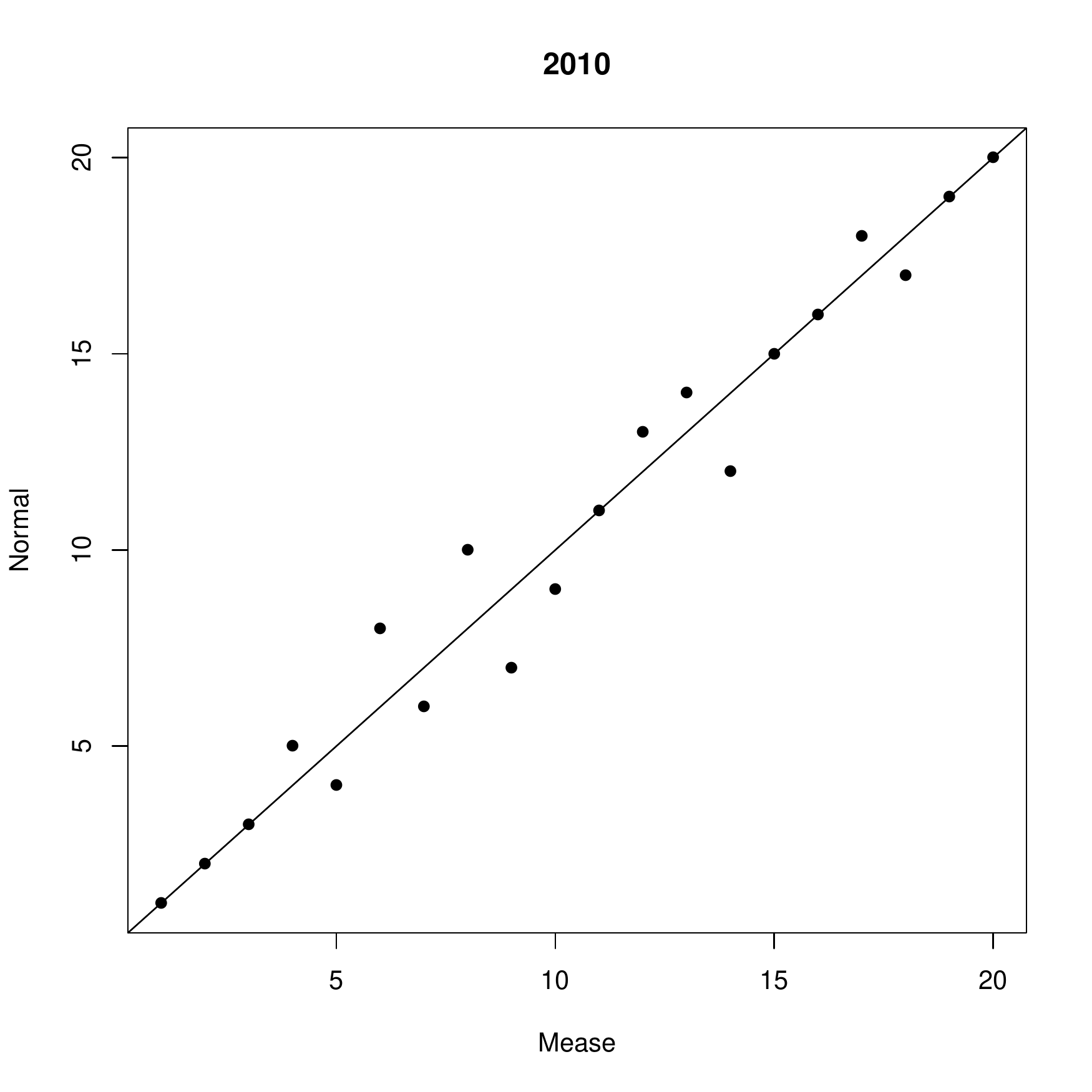}
\includegraphics[scale=.35]{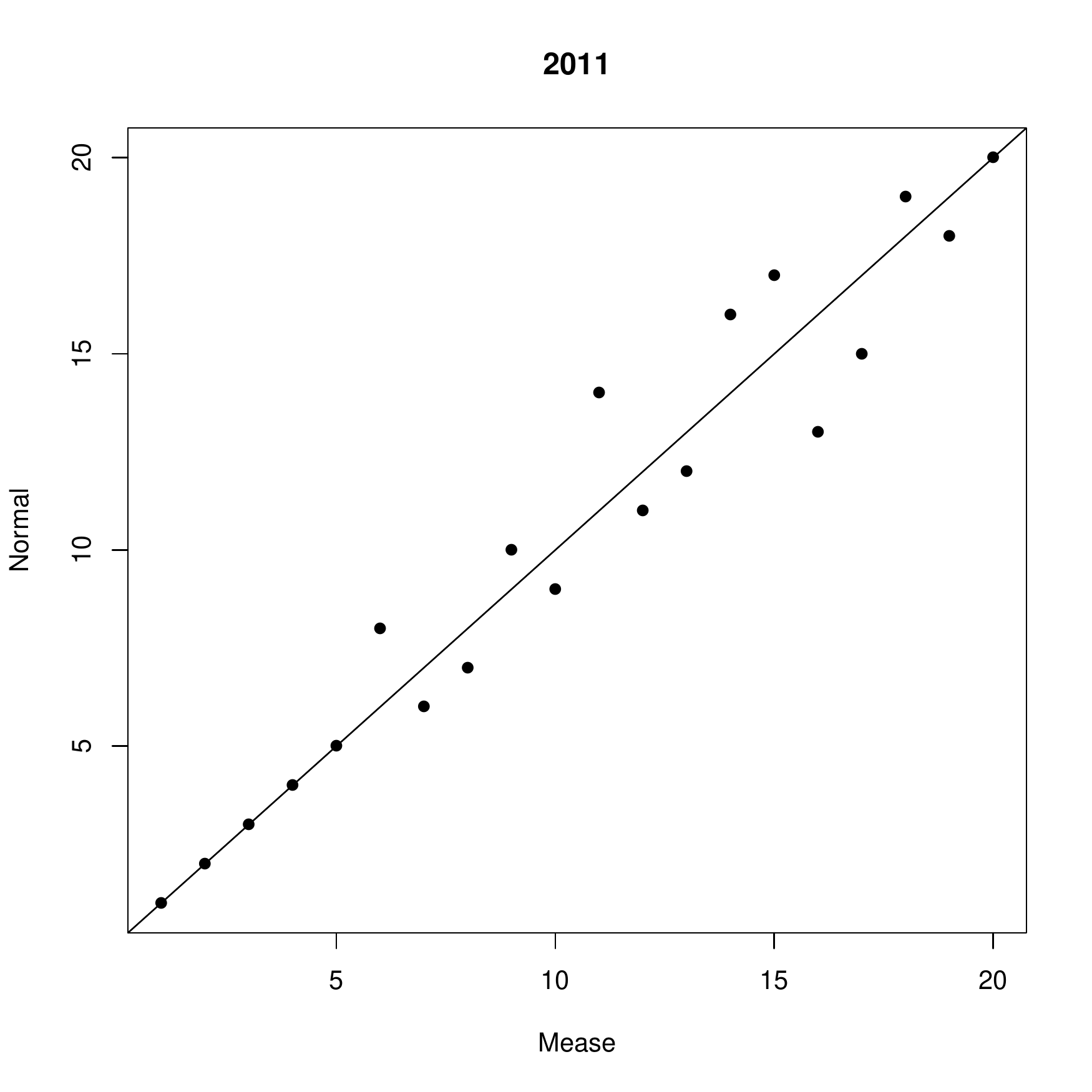}
\end{figure}

\citet{annis} express concern about potential sensitivity of team rankings to the penalty function chosen by \citet{mease}. In Section \ref{ssec:fe.p.0}, we demonstrated that the penalty function (\ref{m2}) corresponds to the distribution of the random team-effects. Figure \ref{plot:dist_comp} compares the normal distribution to the distribution implied by the penalty function of \citet{mease}. Figure \ref{plot:dist_comp2} compares the rankings from PQL.P.0 and the model by \citet{mease}. We summarize the changes to the top five teams when moving from PQL.P.0 to \citet{mease}. In 2008, Texas Tech moves from 5 to 4 and Florida moves from 4 to 5. In 2009, Texas moves from 2 to 3 and Cincinnati moves from 3 to 2. In 2010, Stanford moves from 5 to 4 and Oklahoma moves from 4 to 5. There are no changes in the top 5 in 2011, but there are several changes farther down in the top 20.

The distribution (\ref{m2}) used by \citet{mease} is fixed and does not depend upon estimated parameters. It is very similar to a $N(\bds{0},0.815*\bds{I})$ distribution. In this sense, Mease's model may be approximated using PQL.P.0 by restricting $\sigma^2_t=.815$. To address the point made by \citet{annis}, we fit PQL.P.0 using different fixed values of $\sigma^2_t$. At one extreme, consider the case $\sigma^2_t=0$. This implies that all of the teams are of equal strength, that that the chance of a team winning any given game is 50\%, and that the teams should be ranked by their number of wins minus their number of losses. To explore this possibility, we restricted $\sigma^2_t=0.0001$ in our R program (which requires a positive value of $\sigma^2_t$) and found roughly what we expected in the left column of Table \ref{tab:fake}. Notice how Arkansas St.~made it up to 11 under this scenario. For the other extreme, we restricted $\sigma^2_t=100$. Notice how 12 of the top 15 teams are from either the SEC or Big XII, including 7-5 Auburn and Texas, and 6-6 Texas A\&M. Note that Arkansas' two losses were to LSU and Alabama, and that Alabama's only loss was to LSU. 

Our model provides an advantage over that of \citet{mease} by estimating $\sigma^2_t$ rather than fixing it at an arbitrary value. This discussion also provides at least a partial explanation for the changes in ranking due to changes of integral approximation. The downward bias in the PQL estimates of $\sigma^2_t$ \citep{breslow95} has the same effect as modifying the assumed distribution of the random effects. Across the years, the PQL.P.0 estimates of $\sigma^2_t$ tend to be around $0.55$, while the FE.P.0 estimates tend to be closer to $0.8$. This seems to explain why Mease's model tends to agree more closely with FE.P.0 instead of PQL.P.0, despite relying on a PQL approximation.

% Table generated by Excel2LaTeX from sheet '100'
% Table generated by Excel2LaTeX from sheet '100'
\begin{table}[htbp]
  \centering
  \caption{Rankings under extreme values of $\sigma^2_t$.}
    \begin{tabular}{rlllrrlll}
    \addlinespace
    \toprule
   &\multicolumn{2}{c}{$\sigma^2_t=0.0001$}&\phantom{abc}&&\multicolumn{2}{c}{$\sigma^2_t=100$}\\    
\cmidrule{1-4}
\cmidrule{6-9}  
    Rank  & Team  & W     & L     &       & Rank  & Team  & W     & L \\
    \midrule
    1     & LSU   & 13    & 0     &       & 1     & LSU   & 13    & 0 \\
    2     & Alabama & 11    & 1     &       & 2     & Alabama & 11    & 1 \\
    3     & Boise St. & 11    & 1     &       & 3     & Arkansas & 10    & 2 \\
    4     & Houston & 12    & 1     &       & 4     & Oklahoma St. & 11    & 1 \\
    5     & Oklahoma St. & 11    & 1     &       & 5     & Kansas St. & 10    & 2 \\
    6     & Stanford & 11    & 1     &       & 6     & South Carolina & 10    & 2 \\
    7     & Oregon & 11    & 2     &       & 7     & Baylor & 9     & 3 \\
    8     & Southern Miss. & 11    & 2     &       & 8     & Oklahoma & 9     & 3 \\
    9     & Virginia Tech & 11    & 2     &       & 9     & Stanford & 11    & 1 \\
    10    & Arkansas & 10    & 2     &       & 10    & Oregon & 11    & 2 \\
    11    & Arkansas St. & 10    & 2     &       & 11    & Georgia & 10    & 3 \\
    12    & Clemson & 10    & 3     &       & 12    & Boise St. & 11    & 1 \\
    13    & Georgia & 10    & 3     &       & 13    & Auburn & 7     & 5 \\
    14    & Kansas St. & 10    & 2     &       & 14    & Texas & 7     & 5 \\
    15    & Michigan & 10    & 2     &       & 15    & Texas A\&M & 6     & 6 \\
    \bottomrule
    \end{tabular}%
  \label{tab:fake}%
\end{table}%

\subsection{Modeling FCS Teams}
We consider three different approaches to handling FBS games against FCS teams, using a fully exponential approximation with a probit link. FE.P.0 uses the same approach as \citet{mease}, discarding all FCS games that did not involve an FBS opponent and consolidating all FCS teams into a single ``team'' in the population of FBS teams. FE.P.1 models a separate FCS population, using a pooled estimate of the FBS and FCS population variances. Finally, FE.P.2 models separate populations, estimating a different variance for each population. The results appear in the tables in Appendix \ref{sec:tables}, and are also compared to Mease's model. The reference lines with slope 1 and intercept 0 in the scatter-plots in Appendix \ref{sec:tables} illustrate the difference in estimated team-effect variances between the models. For example, the plot in position [1,2] in the matrix of plots in Figure \ref{plot:2008_c_2} shows the ratings falling below the reference line. This indicates that, in 2008, the variance of the FBS team-effects from FE.P.2 is less than that of FE.P.1.

\begin{figure}
\caption{Distributions of team ratings from the 2011 FE.P.1 model. The dashed line corresponds to FCS teams, the solid line to FBS teams.}
\label{plot:2011_distributions}
\centering
\includegraphics[scale=.4]{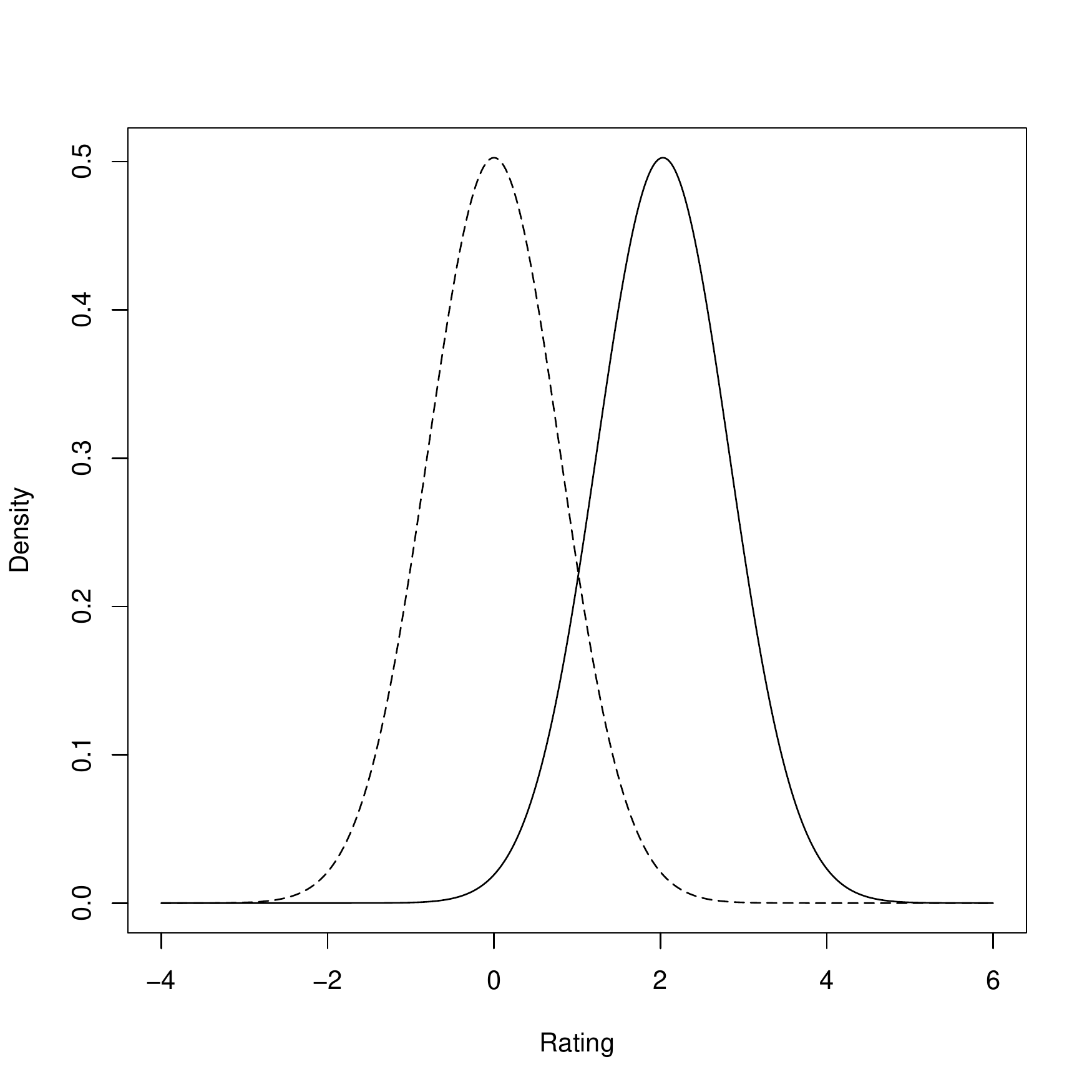}
\end{figure}

Table \ref{tab:sigmas} shows the estimated pooled variance ($\sigma^2_t$) from FE.P.1 and the FBS variance ($\sigma^2_1$) and FCS variance ($\sigma^2_2$) from FE.P.2. Figure \ref{plot:2011_distributions} plots the distributions of the team ratings from FE.P.1 2011. This plot illustrates the difference in means for the two populations, as well as the overlap between the distributions. This difference, obtained via the estimate $\hat{\beta}=2.03$, indicates that the probability of a randomly selected FBS team defeating a randomly selected FCS team in 2011 is $\Phi(2.03)\approx 0.979$. FE.P.0 ranks Alabama 2 in 2011, while FE.P.1 and FE.P.2 pick Oklahoma St. In 2008, FE.P.2 ranks Florida 4, while FE.P.1 and FE.P.0 rank Texas Tech 4. There are other changes lower in the rankings as well. 

FE.P.2 provides greater flexibility than FE.P.1 by estimating separate variance components. Likewise, FE.P.2 makes use of more information than FE.P.0, by considering the outcome of games between pairs FCS teams as well as modeling the FBS games against specific FCS teams rather than a generic FCS ``team.'' We prefer FE.P.2; however, the approach used by FE.P.0 is reasonable as well. As we have seen in previous sections, the choice between two reasonable assumptions may lead to different rankings.

\begin{table}[htbp]
  \centering
  \caption{Estimates for $\sigma^2_t$ from FE.P.1, and $\sigma^2_1$ and $\sigma^2_2$ from FE.P.2}
    \begin{tabular}{rrrrr}
    \addlinespace
    \toprule
    Year  & $\sigma^2_t$&    & $\sigma^2_1$&$\sigma^2_2$ \\
    \midrule
    2008  & 0.75 &&0.65&	0.87\\
    2009  & 0.82 && 0.82&	0.82 \\
    2010  & 0.71 &&0.80&0.60 \\
    2011  & 0.63 &&0.70&0.55 \\
    \bottomrule
    \end{tabular}%
  \label{tab:sigmas}%
\end{table}%

\subsection{ML vs. REML}
For now, we will only consider the sensitivity to the choice of ML versus REML estimation when using PQL in SAS. Under the model PQL.P.1, none of the top 16 teams from 2008-2011 differ between the ML and the REML rankings (not shown). This is not surprising since our model includes only one fixed effect, the FCS effect, and around 240 different levels of the random effect, corresponding to the FBS and FCS teams. Unlike a one-way random effects model, it is not clear what the expected downward bias in the ML estimates of $\sigma^2_t$ should be in this multi-membership random effects setting. However, we found the estimates from two methods to be nearly identical. We include estimates from the two methods in Table \ref{tab:reml} so that the difference between the ML and REML estimates may be compared to the differences resulting from different integral approximations. Of course, the difference between ML and REML estimates would grow if additional fixed effects were added to the model.

\begin{table}[htbp]
  \centering
  \caption{Estimates for $\sigma^2_t$ from PQL.P.1}
    \begin{tabular}{rrr}
    \addlinespace
    \toprule
    Year  & ML    & REML \\
    \midrule
    2008  & 0.4763 & 0.4768 \\
    2009  & 0.5077 & 0.5087 \\
    2010  & 0.4405 & 0.4415 \\
    2011  & 0.4206 & 0.4216 \\
    \bottomrule
    \end{tabular}%
  \label{tab:reml}%
\end{table}%

\section{Conclusion}
We have proposed a generalization of the model developed by \citet{mease} and tested our model's rankings for sensitivity to several modeling choices. Ideally, this type of sensitivity analysis should be performed whenever a generalized linear mixed model is used. The sensitivity will likely depend on the random effects structure of different models. The downward bias in PQL estimates of variance components has been well documented \citep{breslow95}, but not as much attention has been given to differences in EBLUP orderings resulting from using different orders of integral approximation. 

\citet{harville03} discusses seven criteria for an appropriate ranking-model: accuracy, appropriateness, impartiality, unobtrusiveness, nondisruptiveness, verifiability, and comprehensibility. For practical purposes, the choice between Mease's model  and our model FE.P.2 represents a trade-off between comprehensibility and accuracy. The computational effort required to obtain the rankings from FE.P.2 is much greater than that required for Mease's model. Furthermore, the lack of a closed-form objective function obfuscates the relationship between the data and the rankings. However, by modeling a separate FCS population and using a fully exponential Laplace approximation, FE.P.2 makes use of additional data and provides the capacity to accurately estimate the FBS population variance, whereas Mease relies on a fixed population variance.

The sensitivity of the EBLUPs to different methods of approximating the marginal likelihood may be of interest in other settings, including in the use of value-added models for teacher assessment. Value-added models evaluate teachers based on the performance of their students. When the measures of student performance are categorical \citep{broatch10}, the analysis of the sensitivity to the choice of approximation of the marginal likelihood that we discuss in this paper may be relevant. 

Without changing the mean or random effects structures, our rankings shifted with different choices of modeling assumptions. The resulting changes in team ratings are small relative to the standard errors of the ratings, but could have implications for which teams are assigned to which bowls. Bowls are assigned based on the point estimates of the team rankings: an undefeated, third ranked team would probably take little consolation in being told that their rating is not significantly lower than those of the top two teams. 

The large confidence intervals associated with the team ratings suggest that the sensitivity of the rankings to the modeling assumptions is due at
least in part to the limited information available to the model:
around 12 binary outcomes on 240 or so subjects. Given this limited
information, it seems unreasonable to expect these models to be
capable of identifying the two best teams in a division. The limited
accuracy of the models due to their restriction to the use of binary
game outcomes is one of the reasons that lead Stern (2006) to call for
a boycott of the BCS by quantitative analysts. Models for binary game
outcomes may provide a rough guide for the classification of teams, but
the fickle nature of their rankings should be kept in mind.

\appendix
\section{Tables and Graphs}\label{sec:tables}
\begin{figure}
\caption{2008 Division I-A Pre-Bowl Rankings}
\label{plot:2008_d}
\centering
\includegraphics[scale=.5]{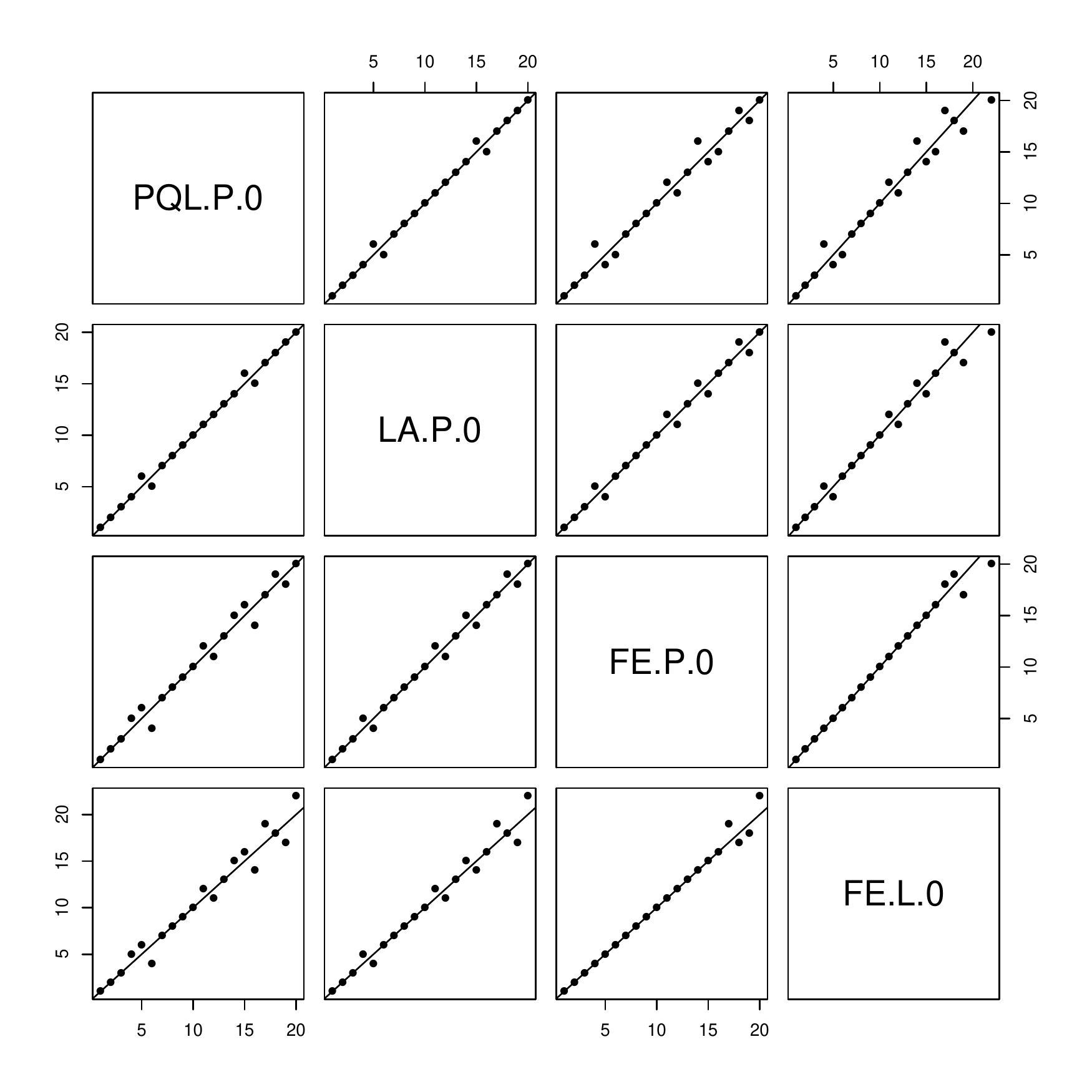}
\end{figure}

\begin{figure}
\caption{2008 Division I-A Pre-Bowl Ratings. Lines have slope 1 and intercept 0.}
\label{plot:2008_c}
\centering
\includegraphics[scale=.5]{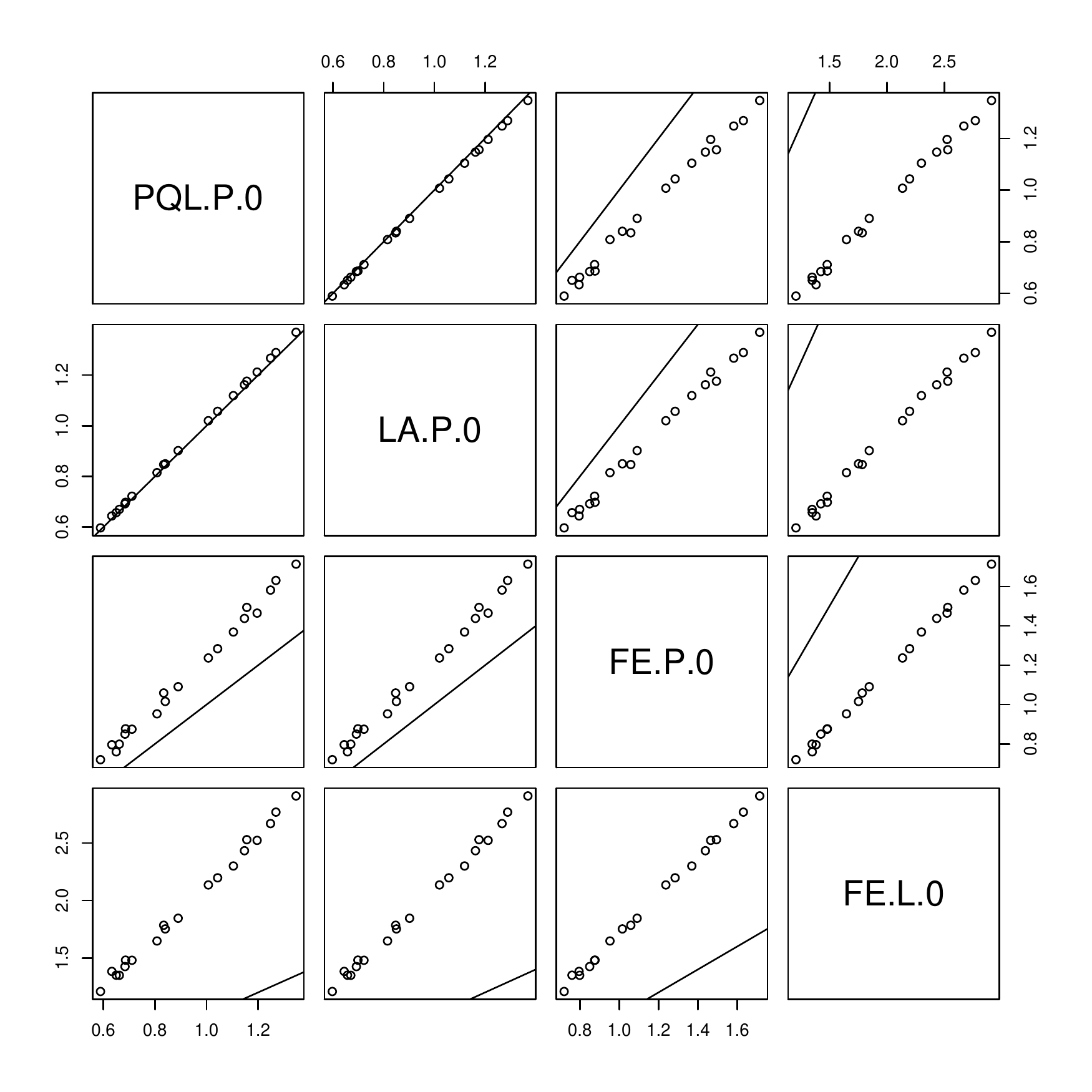}
\end{figure}

Note: For the plots of the rankings (e.g. Figure \ref{plot:2008_d}), the best teams appear in the bottom left corner of each graph, corresponding to rankings near 1. By contrast, the best teams in the plots of the ratings appear in the top right corner, corresponding to larger ratings (e.g. Figure \ref{plot:2008_c}).
\begin{table}[htbp]
  \centering
\caption{2008 Division I-A Pre-Bowl Rankings and Ratings, sorted by FE.P.0}
\label{tab:2008}   
 \resizebox{5in}{!}{
    \begin{tabular}{@{}clrrrrrrrrr@{}}
    \addlinespace
    \toprule   
BCS& Team & &PQL.P.0&  &LA.P.0&&FE.P.0 && &FE.L.0\\
    \midrule
    1     & Oklahoma & 1     & 1.347 & 1     & 1.369 & 1     & 1.714 &       & 1     & 2.911 \\
    6     & Utah  & 2     & 1.269 & 2     & 1.289 & 2     & 1.631 &       & 2     & 2.770 \\
    3     & Texas & 3     & 1.248 & 3     & 1.267 & 3     & 1.582 &       & 3     & 2.670 \\
    7     & Texas Tech & 6     & 1.156 & 5     & 1.176 & 4     & 1.494 &       & 4     & 2.530 \\
    2     & Florida & 4     & 1.196 & 4     & 1.212 & 5     & 1.465 &       & 5     & 2.524 \\
    9     & Boise St. & 5     & 1.147 & 6     & 1.162 & 6     & 1.438 &       & 6     & 2.434 \\
    4     & Alabama & 7     & 1.104 & 7     & 1.119 & 7     & 1.369 &       & 7     & 2.301 \\
    5     & Southern Cal & 8     & 1.043 & 8     & 1.057 & 8     & 1.284 &       & 8     & 2.198 \\
    8     & Penn St. & 9     & 1.007 & 9     & 1.020  & 9     & 1.237 &       & 9     & 2.136 \\
    10    & Ohio St. & 10    & 0.890  & 10    & 0.902 & 10    & 1.091 &       & 10    & 1.846 \\
    11    & TCU   & 12    & 0.834 & 12    & 0.847 & 11    & 1.059 &       & 11    & 1.785 \\
    12    & Cincinnati & 11    & 0.840  & 11    & 0.850  & 12    & 1.016 &       & 12    & 1.753 \\
    22    & Ball St. & 13    & 0.808 & 13    & 0.815 & 13    & 0.953 &       & 13    & 1.648 \\
    13    & Oklahoma St. & 16    & 0.686 & 15    & 0.698 & 14    & 0.877 &       & 14    & 1.481 \\
    15    & Georgia & 14    & 0.711 & 14    & 0.722 & 15    & 0.875 &       & 15    & 1.480 \\
    16    & BYU   & 15    & 0.684 & 16    & 0.692 & 16    & 0.850  &       & 16    & 1.425 \\
    18    & Michigan St. & 17    & 0.662 & 17    & 0.670  & 17    & 0.799 &       & 19    & 1.349 \\
    14    & Georgia Tech & 19    & 0.633 & 19    & 0.644 & 18    & 0.796 &       & 17    & 1.383 \\
    20    & Pittsburgh & 18    & 0.650  & 18    & 0.657 & 19    & 0.760  &       & 18    & 1.350 \\
    17    & Oregon & 20    & 0.589 & 20    & 0.597 & 20    & 0.720  &       & 22    & 1.208 \\
\midrule
&$\sigma^2_t$ &&0.52&  & 0.54&  & 0.76&&&   2.19 \\
\bottomrule
    \end{tabular}%
    }
\end{table}%

\clearpage

\begin{table}[htbp]
  \centering
  \caption{2008 Division I-A Pre-Bowl Rankings and Ratings, sorted by FE.P.2}
   \resizebox{5in}{!}{
    \begin{tabular}{clrrrrrrrr}
    \addlinespace
    \toprule
    BCS   & Team  && FE.P.2 &       & FE.P.1 &       & FE.P.0 &       & Mease  \\
    \midrule
   1     & Oklahoma & 1     & 1.589 & 1     & 1.700 & 1     & 1.714 & 1     & 1.648 \\
    6     & Utah  & 2     & 1.513 & 2     & 1.619 & 2     & 1.631 & 2     & 1.538 \\
    3     & Texas & 3     & 1.469 & 3     & 1.569 & 3     & 1.582 & 3     & 1.512 \\
    2     & Florida & 4     & 1.381 & 5     & 1.459 & 5     & 1.465 & 5     & 1.399 \\
    7     & Texas Tech & 5     & 1.375 & 4     & 1.480 & 4     & 1.494 & 4     & 1.433 \\
    9     & Boise St. & 6     & 1.342 & 6     & 1.422 & 6     & 1.438 & 6     & 1.338 \\
    4     & Alabama & 7     & 1.290 & 7     & 1.367 & 7     & 1.369 & 7     & 1.296 \\
    5     & Southern Cal & 8     & 1.207 & 8     & 1.275 & 8     & 1.284 & 8     & 1.214 \\
    8     & Penn St. & 9     & 1.162 & 9     & 1.228 & 9     & 1.237 & 9     & 1.169 \\
    10    & Ohio St. & 10    & 1.024 & 10    & 1.082 & 10    & 1.091 & 10    & 1.033 \\
    11    & TCU   & 11    & 0.981 & 11    & 1.050 & 11    & 1.059 & 11    & 1.003 \\
    12    & Cincinnati & 12    & 0.956 & 12    & 1.007 & 12    & 1.016 & 12    & 0.965 \\
    22    & Ball St. & 13    & 0.912 & 13    & 0.947 & 13    & 0.953 & 13    & 0.880 \\
    15    & Georgia & 14    & 0.820 & 14    & 0.871 & 15    & 0.875 & 15    & 0.835 \\
    13    & Oklahoma St. & 15    & 0.804 & 15    & 0.866 & 14    & 0.877 & 14    & 0.837 \\
    16    & BYU   & 16    & 0.800 & 16    & 0.848 & 16    & 0.850 & 16    & 0.794 \\
    18    & Michigan St. & 17    & 0.754 & 18    & 0.794 & 17    & 0.799 & 18    & 0.755 \\
    14    & Georgia Tech & 18    & 0.744 & 17    & 0.799 & 18    & 0.796 & 17    & 0.766 \\
    20    & Pittsburgh & 19    & 0.722 & 19    & 0.754 & 19    & 0.760 & 19    & 0.724 \\
    19    & Virginia Tech & 20    & 0.678 & 20    & 0.726 & 21    & 0.719 & 21    & 0.692 \\
    \bottomrule
    \end{tabular}%
    }
  \label{tab:20082}%
\end{table}%
\clearpage
\begin{figure}
\caption{2008 Division I-A Pre-Bowl Rankings}
\label{plot:2008_d_2}
\centering
\includegraphics[scale=.5]{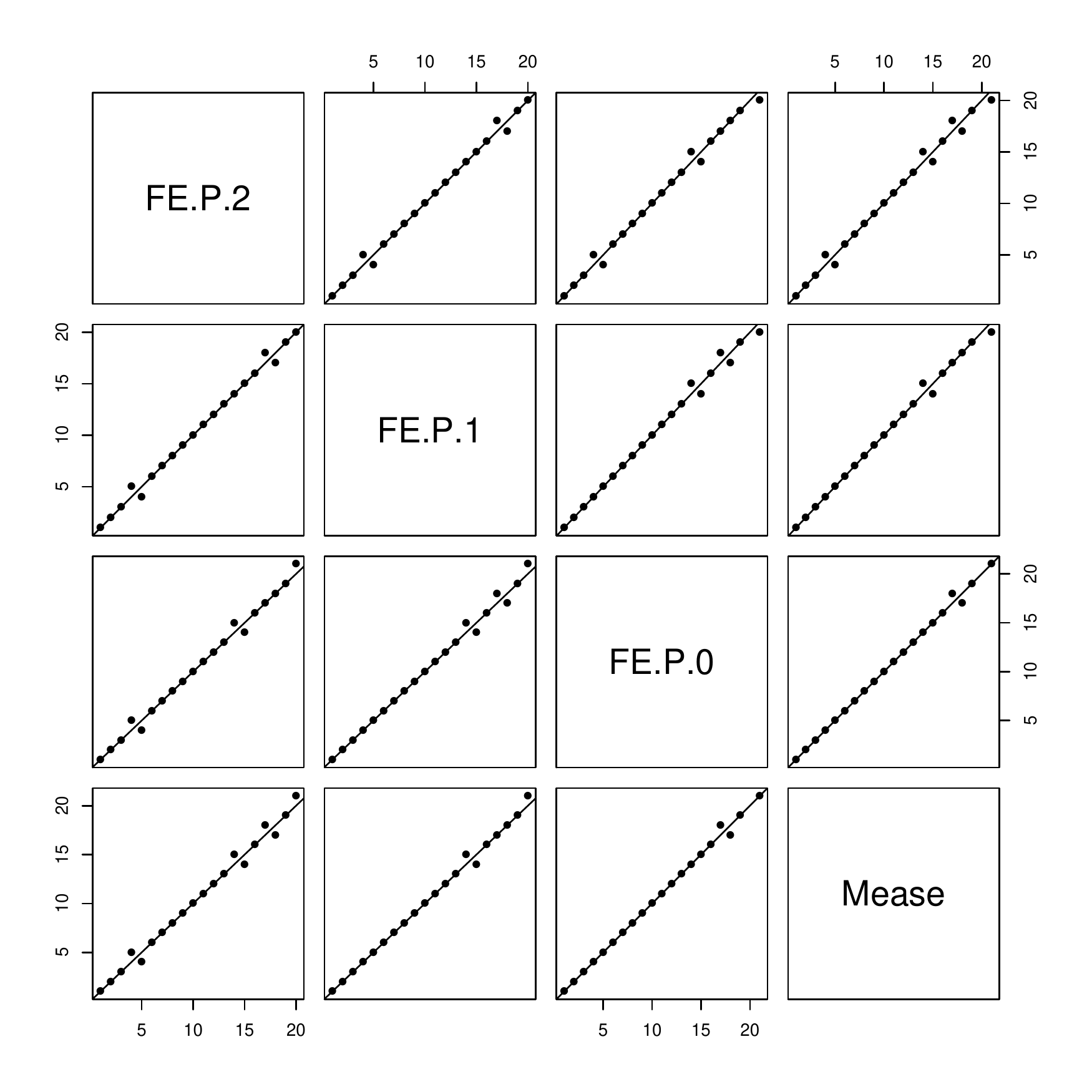}
\end{figure} 

\begin{figure}
\caption{2008 Division I-A Pre-Bowl Ratings}
\label{plot:2008_c_2}
\centering
\includegraphics[scale=.5]{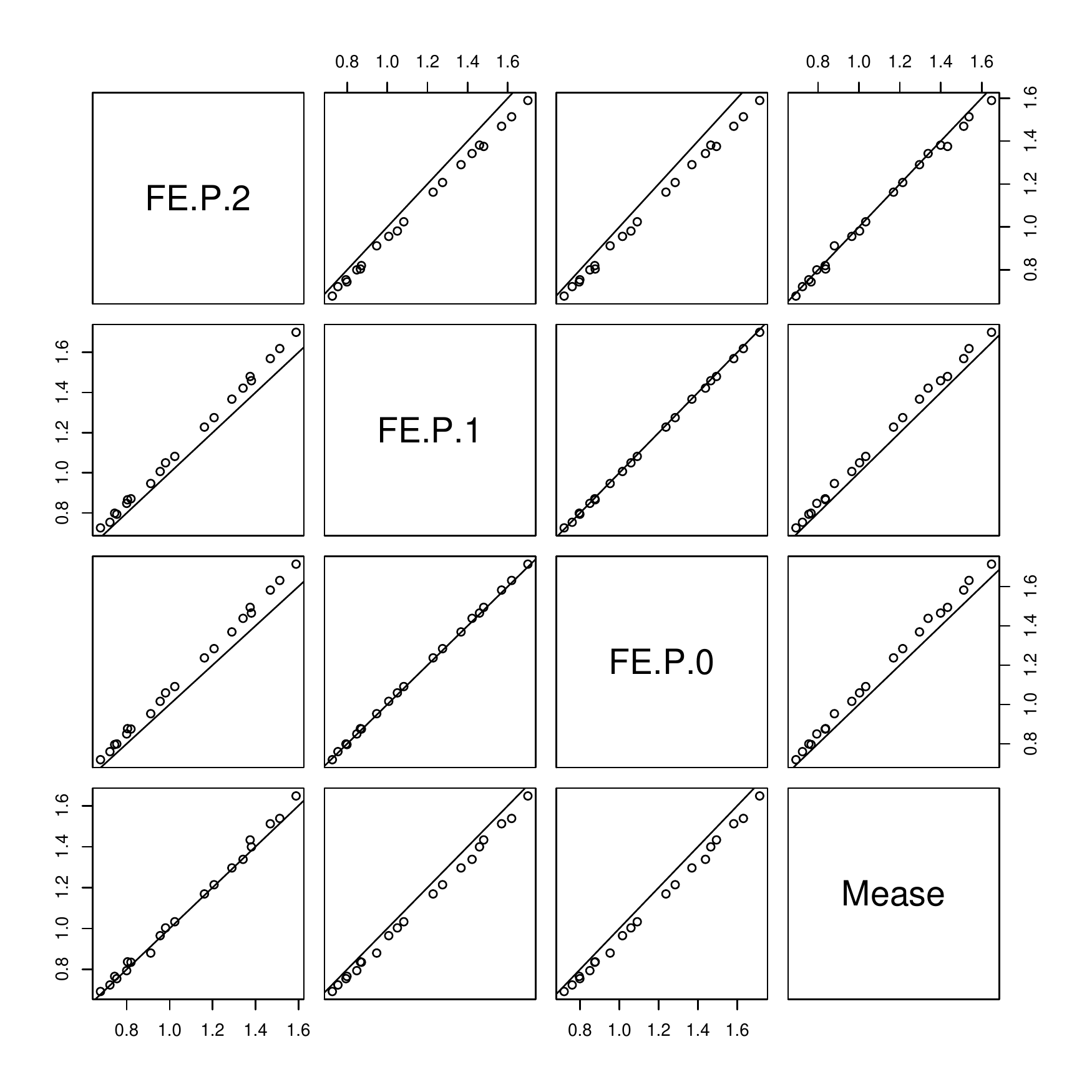}
\end{figure} 

\begin{table}[htbp]
  \centering
 \caption{2009 Division I-A Pre-Bowl Rankings and Ratings, sorted by FE.P.0}
\label{tab:2009}   
 \resizebox{5in}{!}{
    \begin{tabular}{@{}clrrrrrrrrr@{}}
    \addlinespace
    \toprule
BCS& Team & &PQL.P.0&  &LA.P.0&&FE.P.0 && &FE.L.0\\
    \midrule
    1     & Alabama & 1     & 1.592 & 1     & 1.618 & 1     & 2.064 &       & 1     & 3.543 \\
    3     & Cincinnati & 3     & 1.390  & 3     & 1.412 & 2     & 1.789 &       & 2     & 3.077 \\
    2     & Texas & 2     & 1.413 & 2     & 1.431 & 3     & 1.781 &       & 3     & 3.043 \\
    5     & Florida & 4     & 1.309 & 4     & 1.330  & 4     & 1.675 &       & 4     & 2.855 \\
    4     & TCU   & 5     & 1.262 & 5     & 1.280  & 5     & 1.624 &       & 5     & 2.799 \\
    6     & Boise St. & 6     & 1.256 & 6     & 1.273 & 6     & 1.616 &       & 6     & 2.777 \\
    7     & Oregon & 7     & 1.041 & 7     & 1.056 & 7     & 1.303 &       & 7     & 2.247 \\
    9     & Georgia Tech & 8     & 0.904 & 8     & 0.916 & 8     & 1.112 &       & 8     & 1.948 \\
    10    & Iowa  & 9     & 0.864 & 9     & 0.875 & 9     & 1.051 &       & 9     & 1.835 \\
    8     & Ohio St. & 10    & 0.832 & 10    & 0.841 & 10    & 1.000     &       & 10    & 1.796 \\
    12    & LSU   & 12    & 0.779 & 12    & 0.791 & 11    & 0.997 &       & 11    & 1.718 \\
    11    & Virginia Tech & 11    & 0.786 & 11    & 0.797 & 12    & 0.970  &       & 12    & 1.686 \\
    13    & Penn St. & 13    & 0.760  & 13    & 0.768 & 13    & 0.921 &       & 14    & 1.580 \\
    14    & BYU   & 14    & 0.754 & 14    & 0.763 & 14    & 0.920  &       & 13    & 1.580\\
    16    & West Virginia & 15    & 0.684 & 15    & 0.694 & 15    & 0.844 &       & 15    & 1.477 \\
    15    & Miami (FL) & 16    & 0.681 & 16    & 0.691 & 16    & 0.840  &       & 16    & 1.466 \\
    19    & Oklahoma St. & 17    & 0.654 & 17    & 0.662 & 17    & 0.798 &       & 18    & 1.379 \\
    17    & Pittsburgh & 18    & 0.637 & 18    & 0.645 & 18    & 0.776 &       & 17    & 1.392 \\
    20    & Arizona & 21    & 0.581 & 21    & 0.591 & 19    & 0.746 &       & 19    & 1.306 \\
    18    & Oregon St. & 23    & 0.563 & 22    & 0.574 & 20    & 0.733 &       & 20    & 1.261 \\\midrule
      & $\sigma^2_t$& & 0.57& & 0.59& & 0.85& & &    2.54 \\
\bottomrule
    \end{tabular}%
    }
\end{table}%
\begin{figure}
\caption{2009 Division I-A Pre-Bowl Rankings}
\label{plot:2009_d}
\centering
\includegraphics[scale=.5]{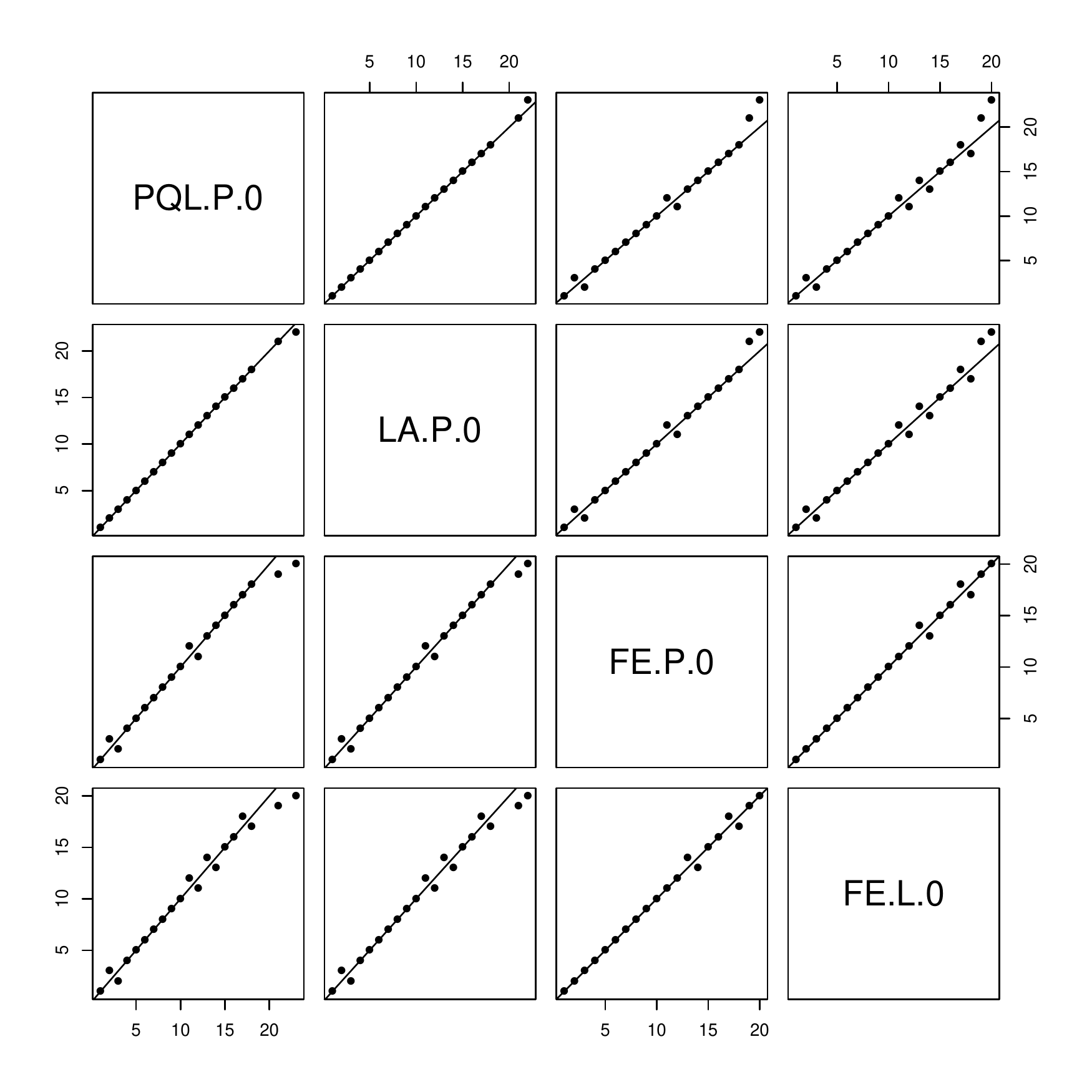}
\end{figure} 

\begin{figure}
\caption{2009 Division I-A Pre-Bowl Ratings}
\label{plot:2009_c}
\centering
\includegraphics[scale=.5]{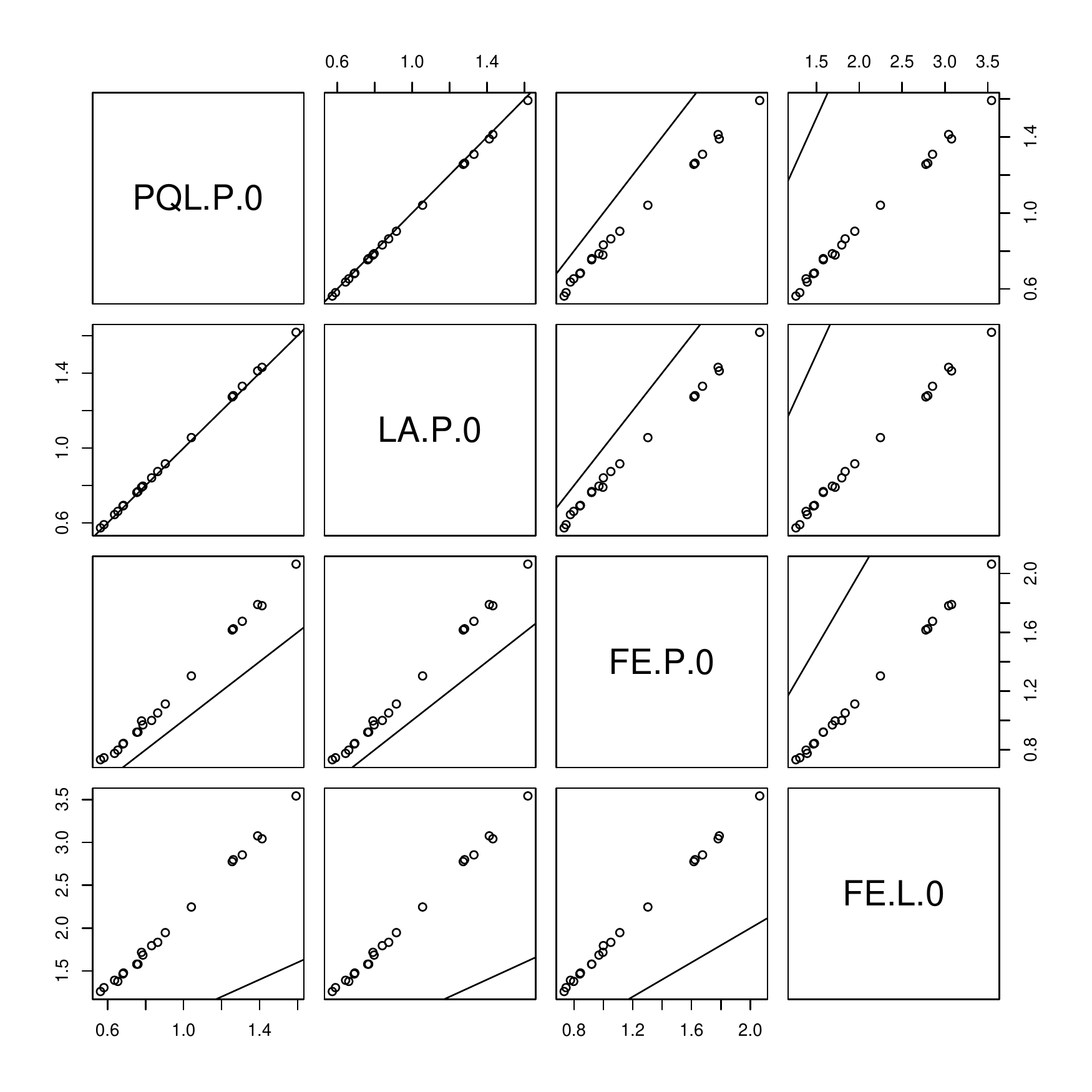}
\end{figure} 

% Table generated by Excel2LaTeX from sheet 'table_2009_2'
\begin{table}[htbp]
  \centering
   \caption{2009 Division I-A Pre-Bowl Rankings and Ratings, sorted by FE.P.2}
   \resizebox{5in}{!}{
    \begin{tabular}{clrrrrrrrr}
    \addlinespace
    \toprule
    BCS   & Team  && FE.P.2 &       & FE.P.1 &       & FE.P.0 &       & Mease  \\
    \midrule
      1     & Alabama & 1     & 2.019 & 1     & 2.016 & 1     & 2.064 & 1     & 1.883 \\
    3     & Cincinnati & 2     & 1.745 & 2     & 1.742 & 2     & 1.789 & 2     & 1.616 \\
    2     & Texas & 3     & 1.742 & 3     & 1.740 & 3     & 1.781 & 3     & 1.609 \\
    5     & Florida & 4     & 1.635 & 4     & 1.633 & 4     & 1.675 & 4     & 1.527 \\
    4     & TCU   & 5     & 1.594 & 5     & 1.592 & 5     & 1.624 & 5     & 1.450 \\
    6     & Boise St. & 6     & 1.576 & 6     & 1.574 & 6     & 1.616 & 6     & 1.440 \\
    7     & Oregon & 7     & 1.271 & 7     & 1.270 & 7     & 1.303 & 7     & 1.191 \\
    9     & Georgia Tech & 8     & 1.103 & 8     & 1.102 & 8     & 1.112 & 8     & 1.019 \\
    10    & Iowa  & 9     & 1.021 & 9     & 1.021 & 9     & 1.051 & 9     & 0.959 \\
    8     & Ohio St. & 10    & 0.977 & 10    & 0.976 & 10    & 1.000 & 10    & 0.914 \\
    11    & Virginia Tech & 11    & 0.965 & 11    & 0.965 & 12    & 0.970 & 12    & 0.890 \\
    12    & LSU   & 12    & 0.961 & 12    & 0.960 & 11    & 0.997 & 11    & 0.908 \\
    13    & Penn St. & 13    & 0.907 & 13    & 0.907 & 13    & 0.921 & 13    & 0.834 \\
    14    & BYU   & 14    & 0.896 & 14    & 0.896 & 14    & 0.920 & 14    & 0.833 \\
    15    & Miami (FL) & 15    & 0.835 & 15    & 0.834 & 16    & 0.840 & 16    & 0.769 \\
    16    & West Virginia & 16    & 0.816 & 16    & 0.816 & 15    & 0.844 & 15    & 0.775 \\
    19    & Oklahoma St. & 17    & 0.768 & 17    & 0.768 & 17    & 0.798 & 17    & 0.732 \\
    17    & Pittsburgh & 18    & 0.746 & 18    & 0.745 & 18    & 0.776 & 18    & 0.714 \\
    20    & Arizona & 19    & 0.717 & 19    & 0.717 & 19    & 0.746 & 19    & 0.678 \\
    18    & Oregon St. & 20    & 0.702 & 20    & 0.701 & 20    & 0.733 & 20    & 0.666 \\
    \bottomrule
    \end{tabular}%
    }
  \label{tab:20092}%
\end{table}%

\clearpage
\begin{figure}
\caption{2009 Division I-A Pre-Bowl Rankings}
\label{plot:2009_d_2}
\centering
\includegraphics[scale=.5]{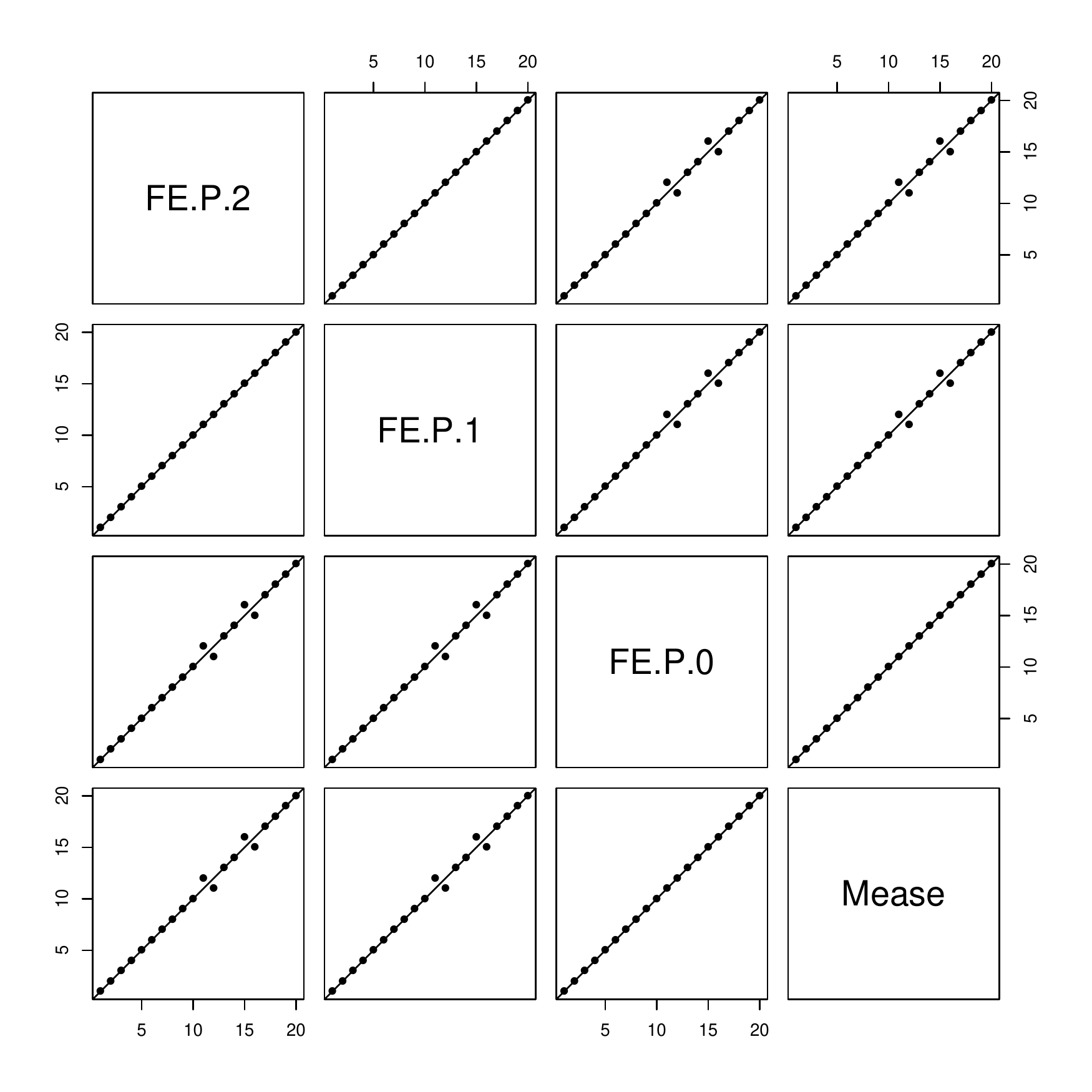}
\end{figure} 

\begin{figure}
\caption{2009 Division I-A Pre-Bowl Ratings}
\label{plot:2009_c_2}
\centering
\includegraphics[scale=.5]{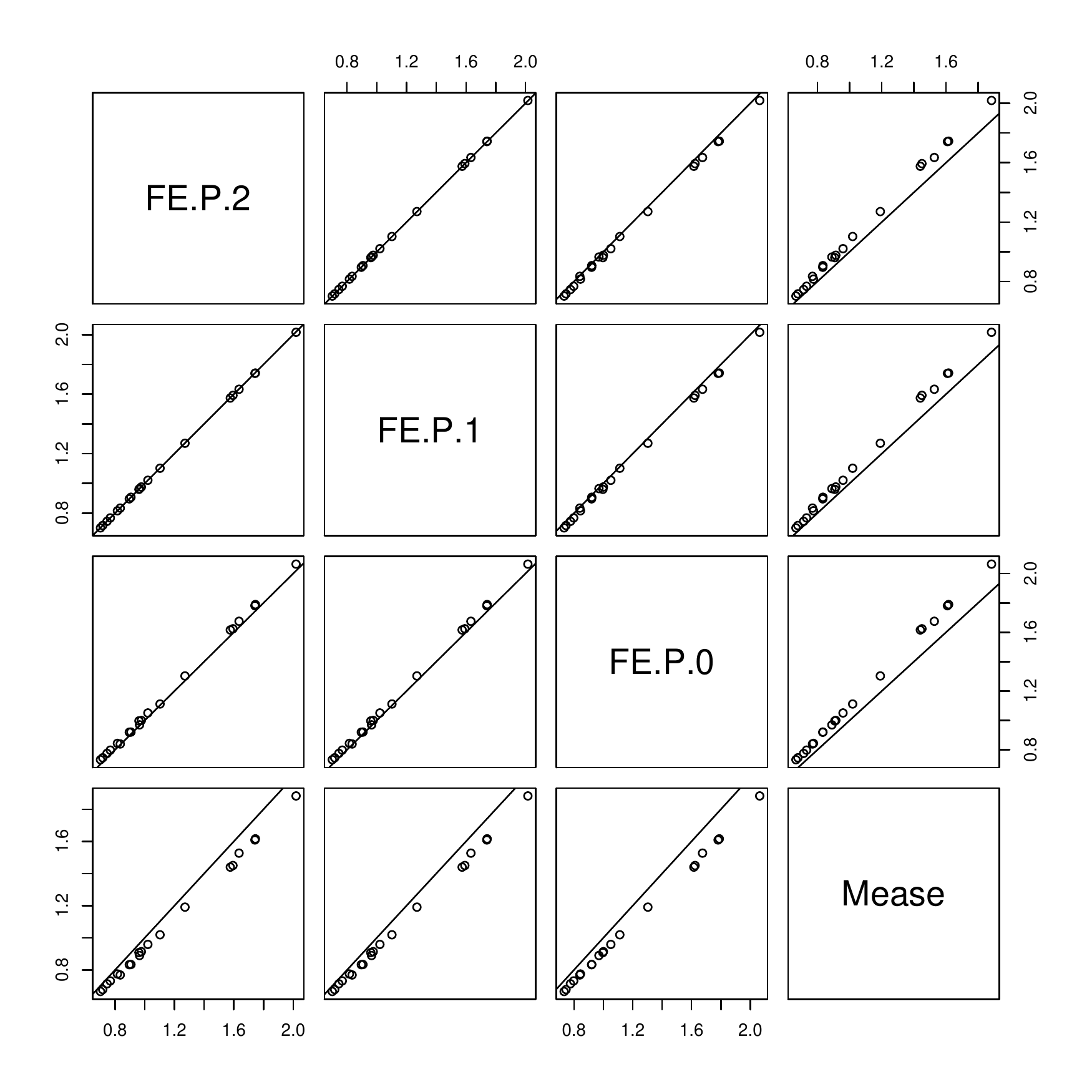}
\end{figure}

\begin{table}[htbp]
  \centering
 \caption{2010 Division I-A Pre-Bowl Rankings and Ratings, sorted by FE.P.0}
\label{tab:2010}   
\resizebox{5in}{!}{
    \begin{tabular}{@{}clrrrrrrrrr@{}}
    \addlinespace
    \toprule
BCS& Team & &PQL.P.0&  &LA.P.0&&FE.P.0 && &FE.L.0\\
    \midrule
      1     & Auburn & 1     & 1.546 & 1     & 1.573 & 1     & 1.996 &       & 1     & 3.423 \\
    2     & Oregon & 2     & 1.326 & 2     & 1.348 & 2     & 1.725 &       & 2     & 2.941 \\
    3     & TCU   & 3     & 1.287 & 3     & 1.307 & 3     & 1.659 &       & 3     & 2.838 \\
    4     & Stanford & 5     & 1.099 & 5     & 1.116 & 4     & 1.400 &       & 4     & 2.370 \\
    7     & Oklahoma & 4     & 1.111 & 4     & 1.128 & 5     & 1.381 &       & 5     & 2.345 \\
    5     & Wisconsin & 8     & 1.073 & 7     & 1.089 & 6     & 1.364 &       & 6     & 2.316 \\
    6     & Ohio St. & 6     & 1.085 & 6     & 1.099 & 7     & 1.347 &       & 9     & 2.277 \\
    8     & Arkansas & 10    & 1.036 & 10    & 1.054 & 8     & 1.331 &       & 8     & 2.288 \\
    9     & Michigan St. & 7     & 1.073 & 8     & 1.088 & 9     & 1.324 &       & 7     & 2.299 \\
    10    & Boise St. & 9     & 1.052 & 9     & 1.066 & 10    & 1.303 &       & 10    & 2.269 \\
    11    & LSU   & 11    & 1.005 & 11    & 1.022 & 11    & 1.280 &       & 11    & 2.185 \\
    15    & Nevada & 12    & 0.977 & 12    & 0.989 & 12    & 1.214 &       & 12    & 2.094 \\
    12    & Missouri & 13    & 0.973 & 13    & 0.988 & 13    & 1.210 &       & 13    & 2.092 \\
    14    & Oklahoma St. & 14    & 0.958 & 14    & 0.972 & 14    & 1.197 &       & 14    & 2.028 \\
    17    & Texas A\&M & 15    & 0.858 & 15    & 0.873 & 15    & 1.095 &       & 15    & 1.857 \\
    16    & Alabama & 16    & 0.795 & 16    & 0.809 & 16    & 1.030 &       & 16    & 1.761 \\
    19    & Utah  & 17    & 0.793 & 17    & 0.804 & 17    & 0.983 &       & 18    & 1.681 \\
    18    & Nebraska & 18    & 0.781 & 18    & 0.794 & 18    & 0.982 &       & 17    & 1.701 \\
    20    & South Carolina & 19    & 0.665 & 19    & 0.675 & 19    & 0.833 &       & 20    & 1.471 \\
    13    & Virginia Tech & 20    & 0.658 & 20    & 0.663 & 20    & 0.749 &       & 19    & 1.631 \\
    \bottomrule
\midrule
      & $\sigma^2_t$ && 0.54& & 0.56 && 0.81 &      && 2.37\\
\bottomrule
    \end{tabular}%
    }
\end{table}%

\begin{figure}
\caption{2010 Division I-A Pre-Bowl Rankings}
\label{plot:2010_d}
\centering
\includegraphics[scale=.5]{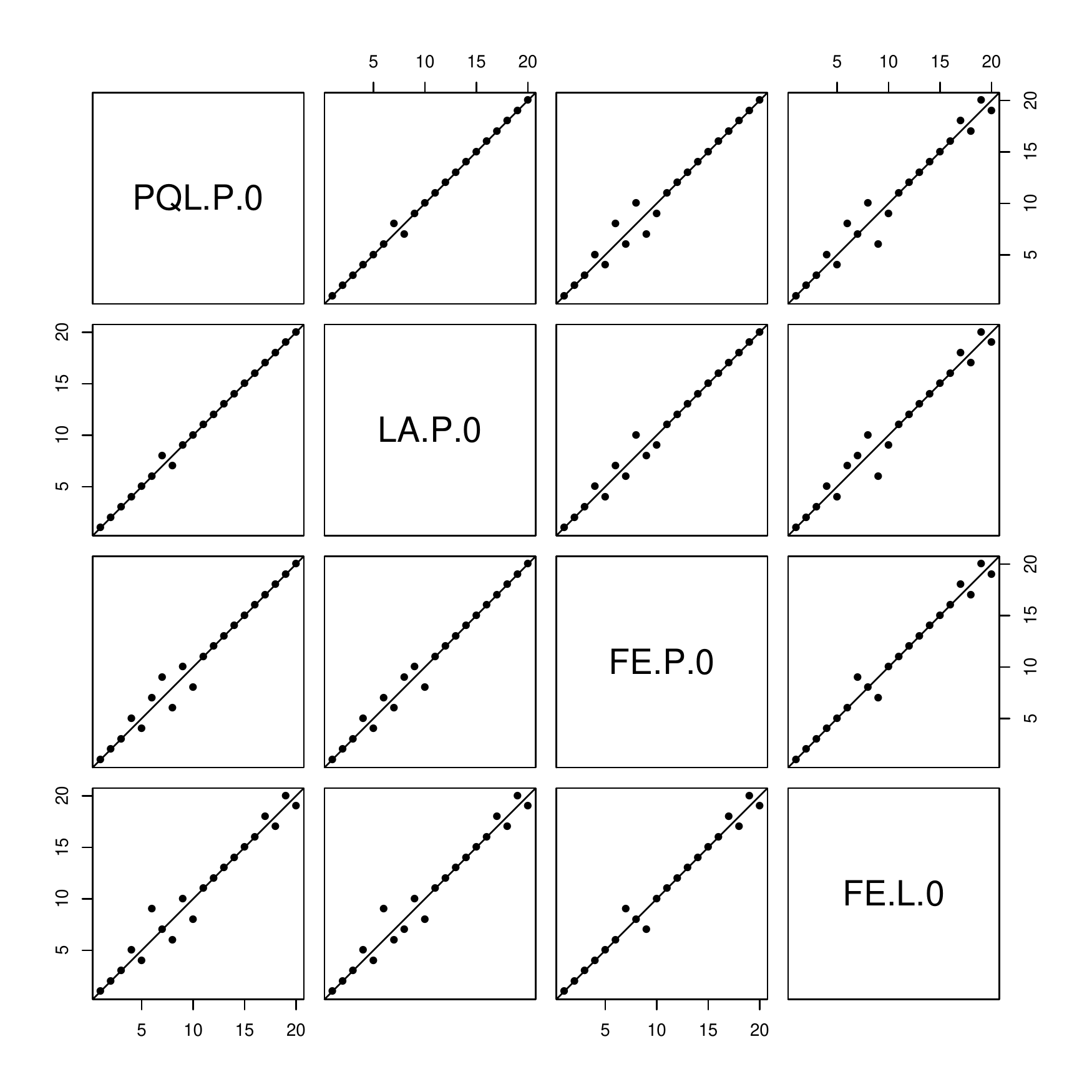}
\end{figure} 

\begin{figure}
\caption{2010 Division I-A Pre-Bowl Ratings}
\label{plot:2010_c}
\centering
\includegraphics[scale=.5]{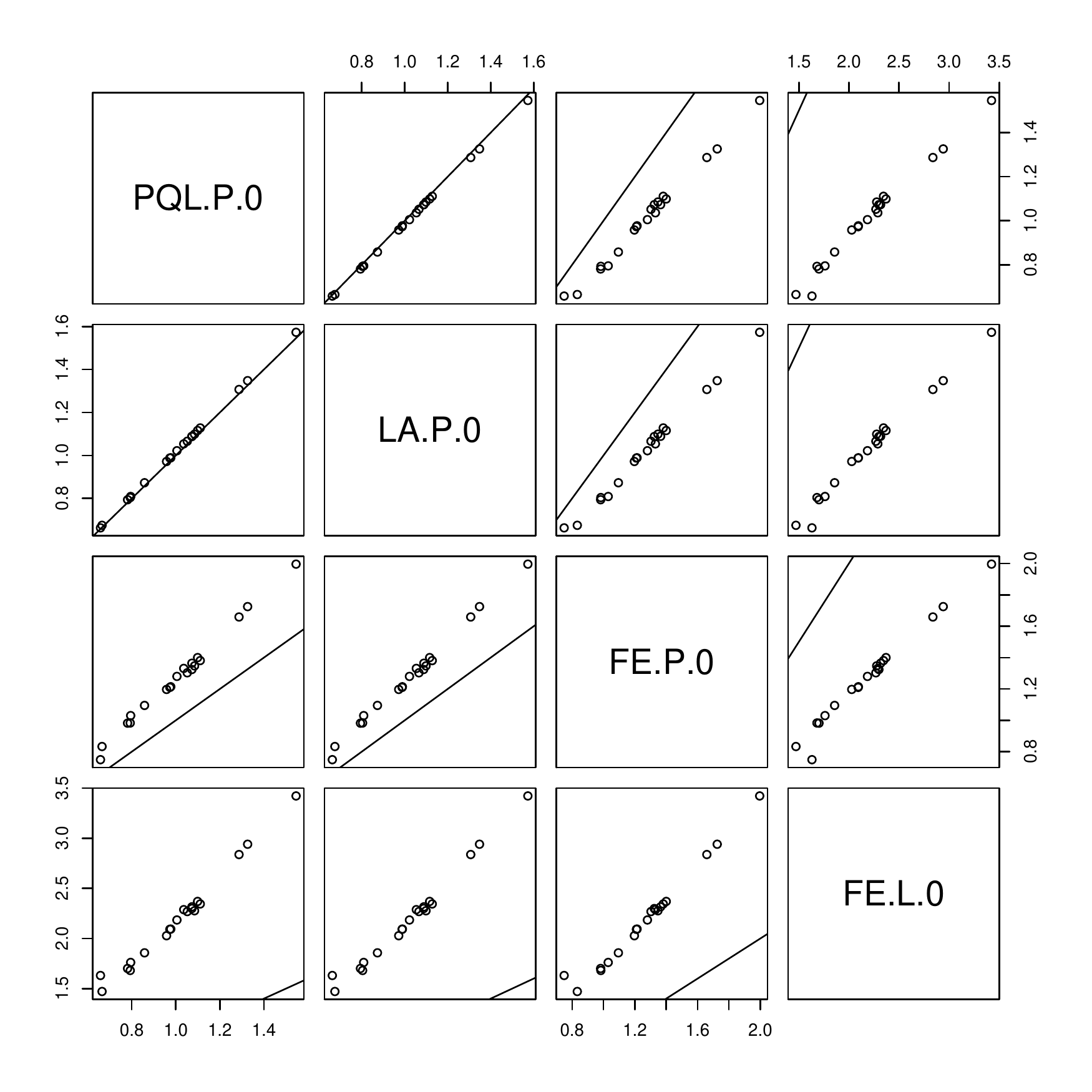}
\end{figure}

% Table generated by Excel2LaTeX from sheet 'table_2010_2'
\begin{table}[htbp]
  \centering
 \caption{2010 Division I-A Pre-Bowl Rankings and Ratings, sorted by FE.P.2}
  \resizebox{5in}{!}{
    \begin{tabular}{clrrrrrrrr}
    \addlinespace
    \toprule
    BCS   & Team  && FE.P.2 &       & FE.P.1 &       & FE.P.0 &       & Mease  \\
    \midrule
     1     & Auburn & 1     & 1.983 & 1     & 1.871 & 1     & 1.996 & 1     & 1.867 \\
    2     & Oregon & 2     & 1.711 & 2     & 1.612 & 2     & 1.725 & 2     & 1.596 \\
    3     & TCU   & 3     & 1.644 & 3     & 1.556 & 3     & 1.659 & 3     & 1.524 \\
    4     & Stanford & 4     & 1.387 & 4     & 1.313 & 4     & 1.400 & 4     & 1.303 \\
    7     & Oklahoma & 5     & 1.371 & 5     & 1.307 & 5     & 1.381 & 5     & 1.292 \\
    5     & Wisconsin & 6     & 1.347 & 6     & 1.278 & 6     & 1.364 & 6     & 1.259 \\
    6     & Ohio St. & 7     & 1.335 & 7     & 1.276 & 7     & 1.347 & 7     & 1.245 \\
    8     & Arkansas & 8     & 1.320 & 9     & 1.245 & 8     & 1.331 & 8     & 1.243 \\
    9     & Michigan St. & 9     & 1.307 & 8     & 1.251 & 9     & 1.324 & 9     & 1.229 \\
    10    & Boise St. & 10    & 1.296 & 10    & 1.241 & 10    & 1.303 & 10    & 1.200 \\
    11    & LSU   & 11    & 1.272 & 11    & 1.205 & 11    & 1.280 & 11    & 1.193 \\
    15    & Nevada & 12    & 1.210 & 12    & 1.159 & 12    & 1.214 & 14    & 1.113 \\
    12    & Missouri & 13    & 1.202 & 13    & 1.145 & 13    & 1.210 & 12    & 1.132 \\
    14    & Oklahoma St. & 14    & 1.190 & 14    & 1.133 & 14    & 1.197 & 13    & 1.114 \\
    17    & Texas A\&M & 15    & 1.091 & 15    & 1.032 & 15    & 1.095 & 15    & 1.023 \\
    16    & Alabama & 16    & 1.021 & 16    & 0.961 & 16    & 1.030 & 16    & 0.962 \\
    18    & Nebraska & 17    & 0.974 & 18    & 0.925 & 18    & 0.982 & 17    & 0.918 \\
    19    & Utah  & 18    & 0.971 & 17    & 0.928 & 17    & 0.983 & 18    & 0.908 \\
    20    & South Carolina & 19    & 0.819 & 19    & 0.778 & 19    & 0.833 & 19    & 0.783 \\
    13    & Virginia Tech & 20    & 0.783 & 20    & 0.768 & 20    & 0.749 & 20    & 0.704 \\
    \bottomrule
    \end{tabular}%
    }
  \label{tab:20102}%
\end{table}%
\clearpage
\begin{figure}
\caption{2010 Division I-A Pre-Bowl Rankings}
\label{plot:2010_d_2}
\centering
\includegraphics[scale=.5]{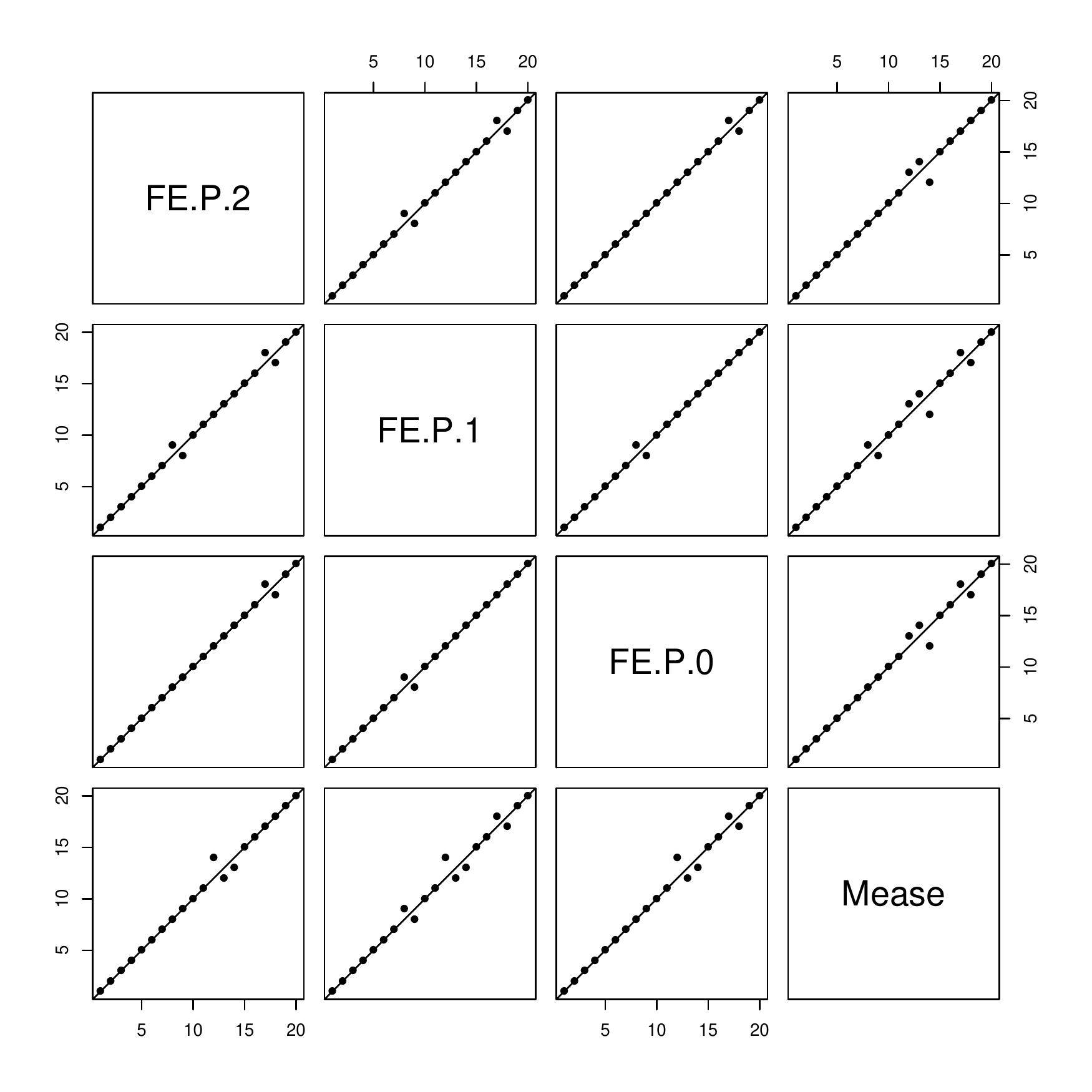}
\end{figure} 

\begin{figure}
\caption{2010 Division I-A Pre-Bowl Ratings}
\label{plot:2010_c_2}
\centering
\includegraphics[scale=.5]{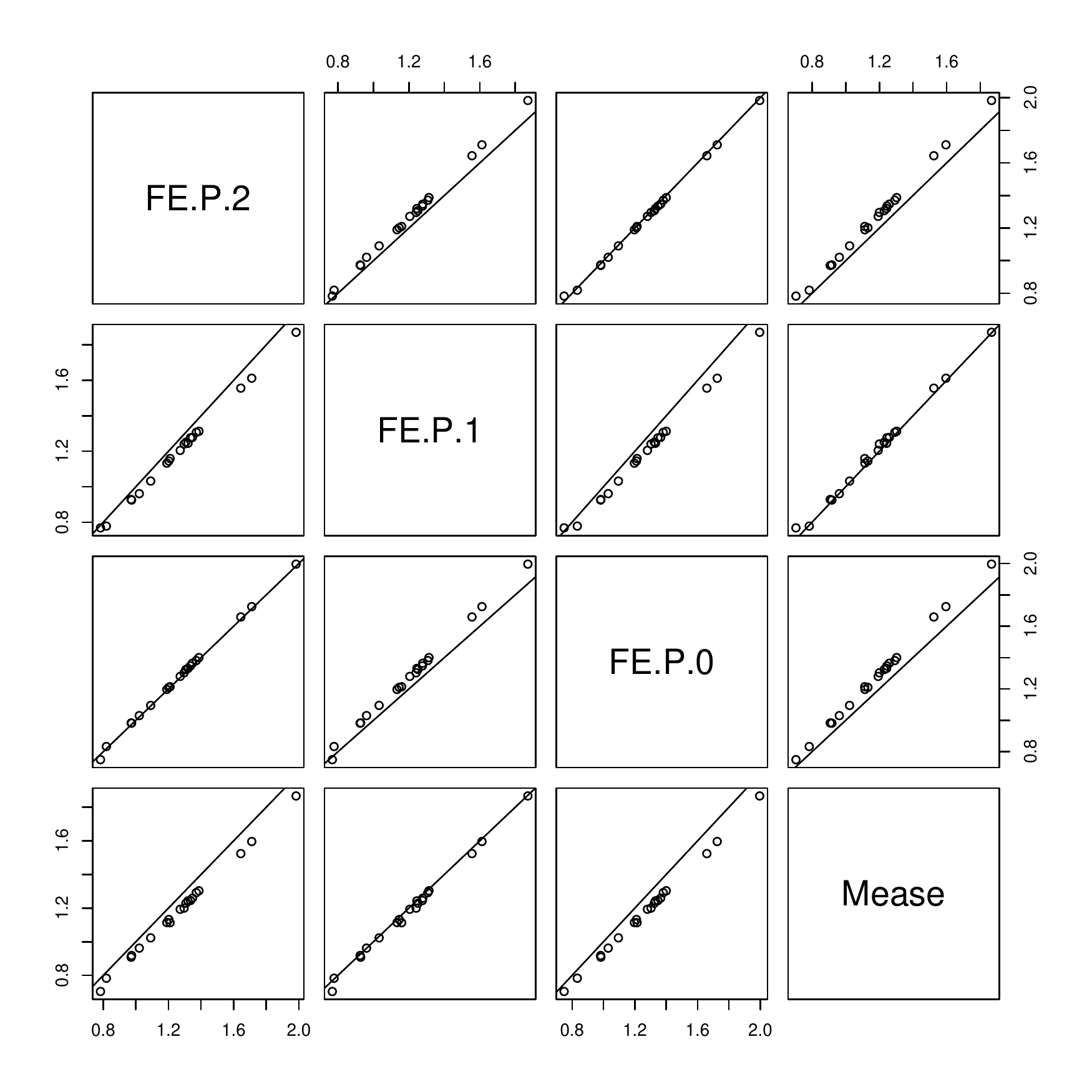}
\end{figure}

\begin{table}[htbp]
  \centering
 \caption{2011 Division I-A Pre-Bowl Rankings and Ratings, sorted by FE.P.0}

\resizebox{5in}{!}{   
    \begin{tabular}{@{}clrrrrrrrrr@{}}
    \addlinespace
    \toprule
BCS& Team &{} &PQL.P.0&{}  &LA.P.0&{}&FE.P.0 &{}& {}&FE.L.0\\
    \midrule
    1     & LSU   & 1     & 1.626 & 1     & 1.651 & 1     & 2.076 &       {}& 1     & 3.544 \\
    2     & Alabama & 3     & 1.233 & 3     & 1.252 & 2     & 1.572 &       {}& 3     & 2.667 \\
    3     & Oklahoma St. & 2     & 1.267 & 2     & 1.286 & 3     & 1.565 &   {}    & 2     & 2.749 \\
    4     & Stanford & 4     & 1.093 & 4     & 1.107 & 4     & 1.347 &       {}& 4     & 2.295 \\
    7     & Boise St. & 5     & 1.069 & 5     & 1.082 & 5     & 1.308 &      {} & 5     & 2.231 \\
    6     & Arkansas & 8     & 0.993 & 7     & 1.010 & 6     & 1.276 &       {}& 6     & 2.168 \\
    8     & Kansas St. & 6     & 1.010 & 6     & 1.026 & 7     & 1.268 &    {}   & 7     & 2.157 \\
    5     & Oregon & 7     & 0.996 & 8     & 1.010 & 8     & 1.233 &      {} & 8     & 2.086 \\
    19    & Houston & 9     & 0.975 & 9     & 0.985 & 9     & 1.166 &      {} & 9     & 2.000 \\
    9     & South Carolina & 10    & 0.930 & 10    & 0.943 & 10    & 1.151 &     {}  & 10    & 1.962 \\
    11    & Virginia Tech & 11    & 0.921 & 11    & 0.931 & 11    & 1.103 &      {} & 13    & 1.870 \\
    14    & Oklahoma & 14    & 0.878 & 13    & 0.892 & 12    & 1.092 &      {} & 11    & 1.903 \\
    probation & Southern Cal & 12    & 0.898 & 12    & 0.909 & 13    & 1.083 &     {}  & 12    & 1.892 \\
    12    & Baylor & 16    & 0.846 & 16    & 0.860 & 14    & 1.069 &      {} & 14    & 1.840 \\
    13    & Michigan & 13    & 0.879 & 14    & 0.888 & 15    & 1.042 &     {}  & 16    & 1.799 \\
    16    & Georgia & 17    & 0.834 & 17    & 0.847 & 16    & 1.042 &      {} & 17    & 1.762 \\
    10    & Wisconsin & 15    & 0.857 & 15    & 0.867 & 17    & 1.031 &     {}  & 15    & 1.819 \\
    18    & TCU   & 18    & 0.764 & 18    & 0.773 & 18    & 0.925 &      {} & 18    & 1.596 \\
    15    & Clemson & 19    & 0.757 & 19    & 0.767 & 19    & 0.916 &    {}   & 19    & 1.590 \\
    17    & Michigan St. & 20    & 0.708 & 20    & 0.717 & 20    & 0.863 &    {}   & 21    & 1.480 \\
\midrule
    {}  & $\sigma^2_t$ &{}& 0.55&{} & 0.57& {}& 0.80&{} &  {}    & 2.33\\
    \bottomrule
    \end{tabular}%
    }
    \label{tab:2011}
\end{table}%

\begin{figure}
\caption{2011 Division I-A Pre-Bowl Rankings}
\label{plot:2011_d}
\centering
\includegraphics[scale=.5]{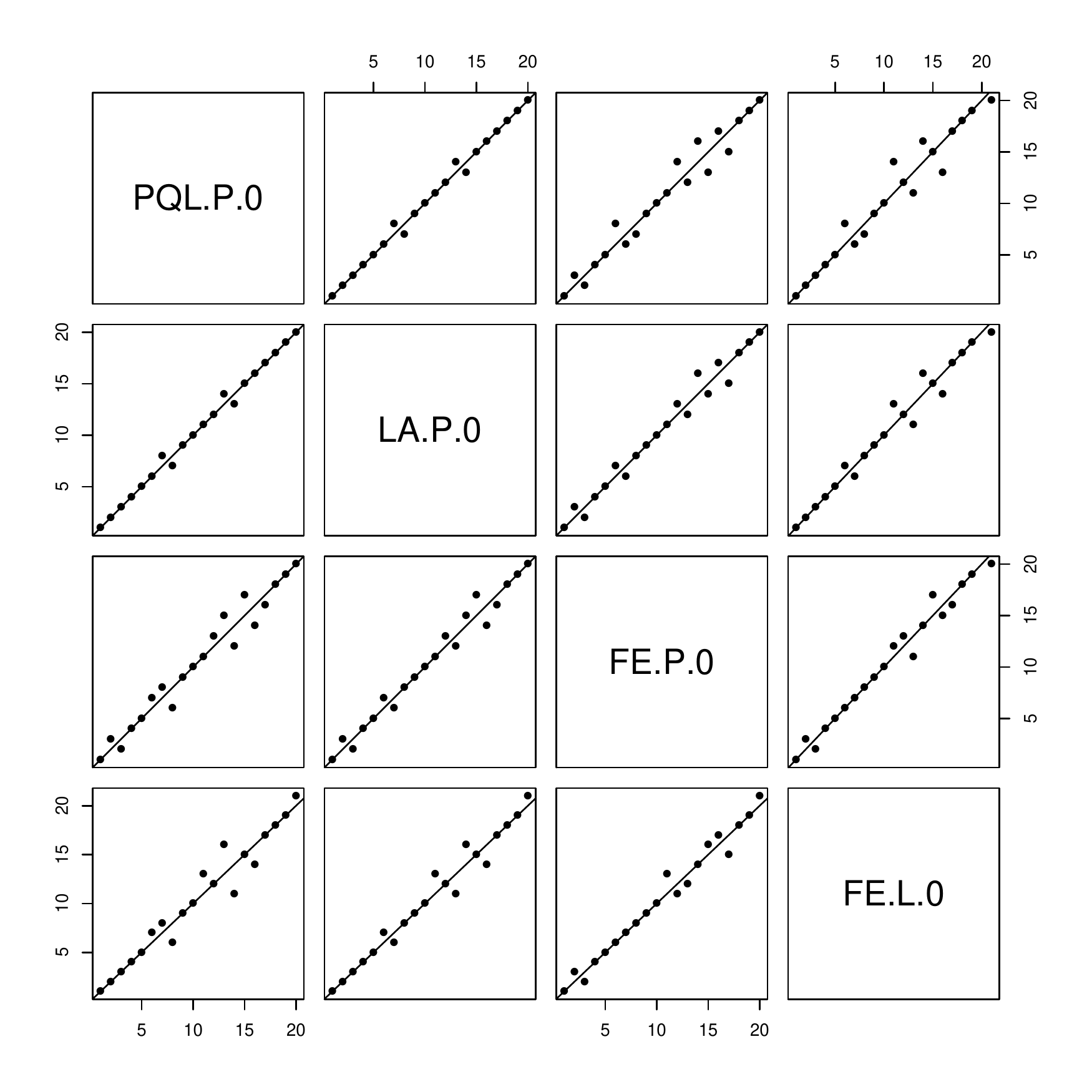}
\end{figure} 

\begin{figure}
\caption{2011 Division I-A Pre-Bowl Ratings}
\label{plot:2011_c}
\centering
\includegraphics[scale=.5]{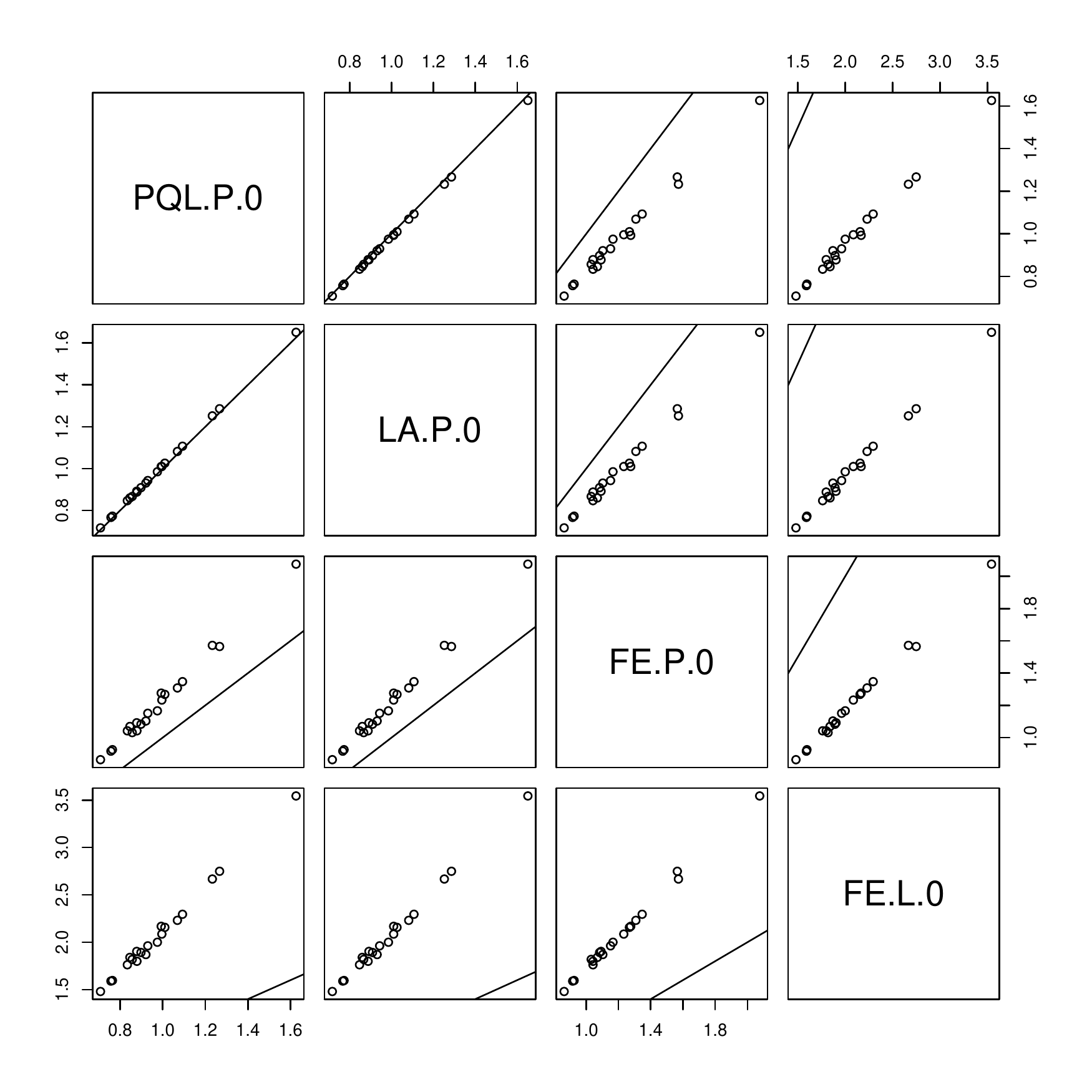}
\end{figure}

% Table generated by Excel2LaTeX from sheet 'table_2011_2'
\begin{table}[htbp]
  \centering
 \caption{2011 Division I-A Pre-Bowl Rankings and Ratings, sorted by FE.P.2}
 \resizebox{5in}{!}{
    \begin{tabular}{clrrrrrrrr}
    \addlinespace
    \toprule
    BCS   & Team  && FE.P.2 &       & FE.P.1 &       & FE.P.0 &       & Mease  \\
				\midrule
   1     & LSU   & 1     & 1.944 & 1     & 1.463 & 1     & 2.076 & 1     & 1.954 \\
    3     & Oklahoma St. & 2     & 1.482 & 2     & 1.152 & 3     & 1.565 & 2     & 1.477 \\
    2     & Alabama & 3     & 1.474 & 3     & 1.108 & 2     & 1.572 & 3     & 1.470 \\
    4     & Stanford & 4     & 1.263 & 4     & 0.986 & 4     & 1.347 & 4     & 1.254 \\
    7     & Boise St. & 5     & 1.243 & 5     & 0.980 & 5     & 1.308 & 5     & 1.209 \\
    8     & Kansas St. & 6     & 1.192 & 6     & 0.906 & 7     & 1.268 & 7     & 1.194 \\
    6     & Arkansas & 7     & 1.185 & 9     & 0.879 & 6     & 1.276 & 6     & 1.196 \\
    5     & Oregon & 8     & 1.151 & 8     & 0.891 & 8     & 1.233 & 8     & 1.154 \\
    19    & Houston & 9     & 1.106 & 7     & 0.897 & 9     & 1.166 & 10    & 1.077 \\
    9     & South Carolina & 10    & 1.077 & 11    & 0.831 & 10    & 1.151 & 9     & 1.080 \\
    11    & Virginia Tech & 11    & 1.044 & 10    & 0.837 & 11    & 1.103 & 12    & 1.027 \\
    14    & Oklahoma & 12    & 1.029 & 14    & 0.789 & 12    & 1.092 & 11    & 1.032 \\
    probation & Southern Cal & 13    & 1.023 & 12    & 0.815 & 13    & 1.083 & 13    & 1.016 \\
    12    & Baylor & 14    & 1.012 & 16    & 0.751 & 14    & 1.069 & 14    & 1.008 \\
    13    & Michigan & 15    & 0.994 & 13    & 0.805 & 15    & 1.042 & 16    & 0.973 \\
    10    & Wisconsin & 16    & 0.979 & 15    & 0.780 & 17    & 1.031 & 17    & 0.961 \\
    16    & Georgia & 17    & 0.971 & 17    & 0.740 & 16    & 1.042 & 15    & 0.978 \\
    18    & TCU   & 18    & 0.877 & 18    & 0.695 & 18    & 0.925 & 19    & 0.853 \\
    15    & Clemson & 19    & 0.862 & 19    & 0.679 & 19    & 0.916 & 18    & 0.859 \\
    17    & Michigan St. & 20    & 0.817 & 20    & 0.638 & 20    & 0.863 & 20    & 0.802 \\
    \bottomrule
    \end{tabular}%
    }
  \label{tab:20112}%
\end{table}%
\clearpage
\begin{figure}
\caption{2011 Division I-A Pre-Bowl Rankings}
\label{plot:2011_d_2}
\centering
\includegraphics[scale=.5]{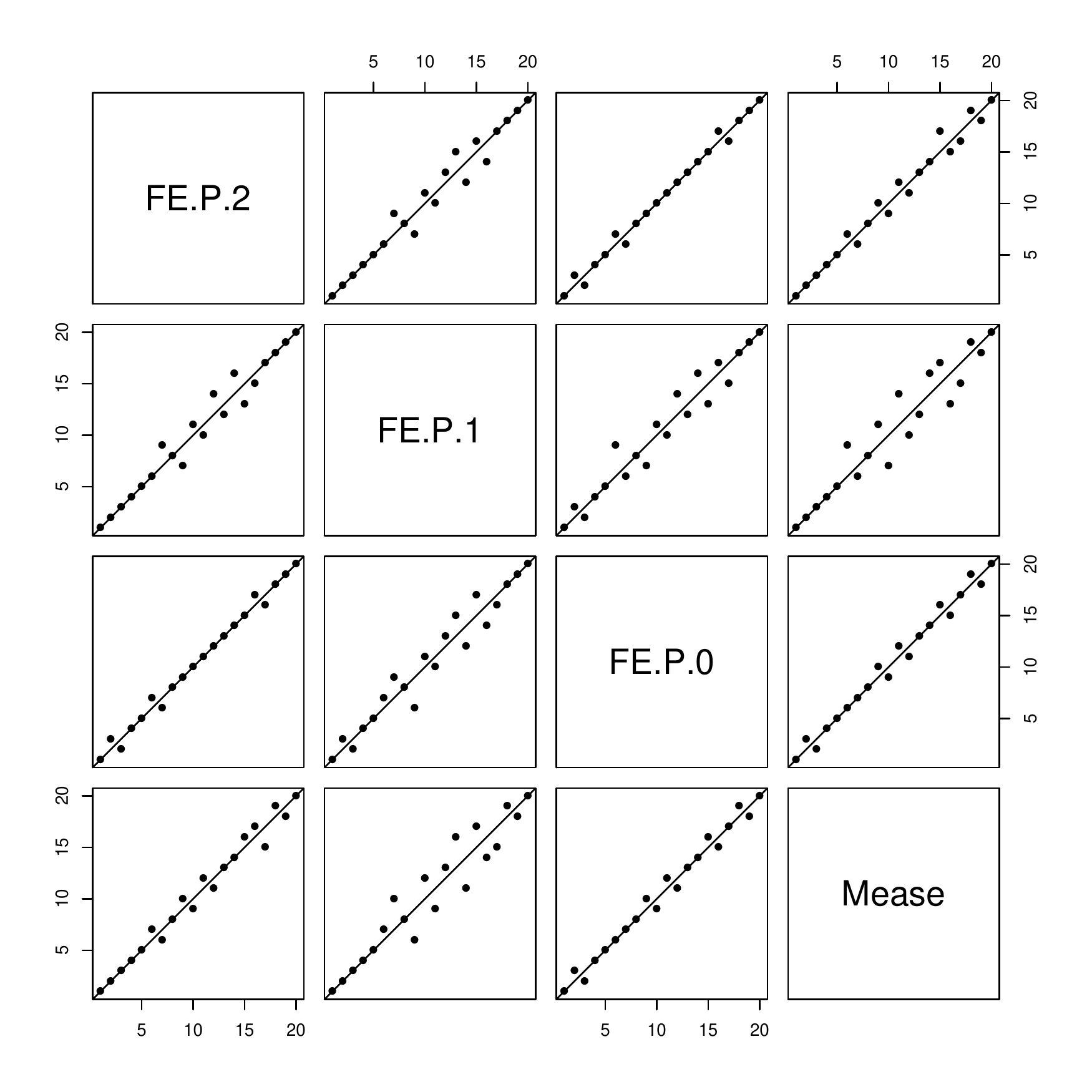}
\end{figure} 

\begin{figure}
\caption{2011 Division I-A Pre-Bowl Ratings}
\label{plot:2011_c_2}
\centering
\includegraphics[scale=.5]{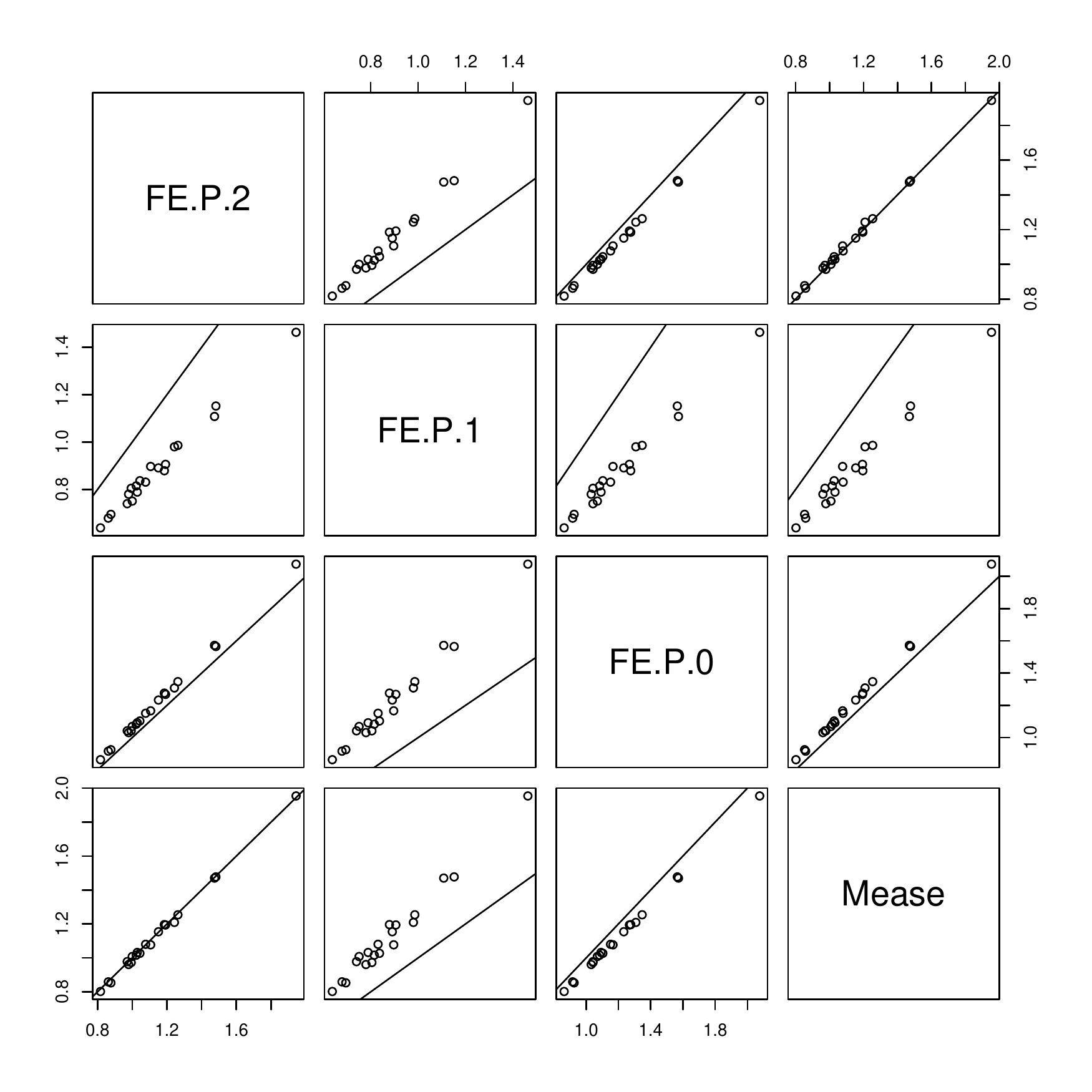}
\end{figure} 

\clearpage

\section{SAS Code}\label{sec:sascode}
The SAS code to produce the 2008 PQL.P.1 rankings appears in Listing \ref{list1}. The EFFECT statement allows us to specify our mutli-membership design matrix by assigning a coefficient of 1, stored in the ``H'' column, to the home team in each game, and a coefficient of -1, stored in the ``A'' column, to the visiting team. The EFFECT statement recognizes that each team appears in both the ``home'' and ``away'' variables, and assigns only one column to each team in the constructed effect we have called ``matchup.'' See Table \ref{tab:data} for the data used by this SAS procedure.
\footnotesize
\newline
\begin{center}
\begin{minipage}{.95\linewidth}
\lstset{language=SAS,frame=single,caption=Example SAS Code, label=list1}
\begin{lstlisting}
PROC GLIMMIX data=d2008 method=mspl; 
	class home away home_win;
	effect matchup = MULTIMEMBER(home away/weight=(H A));
	model home_win (descending)=fcs/noint dist=binary  
		link=probit solution;
	random matchup /solution;
	ods output solutionr=ratings2008;
run;
\end{lstlisting}
\end{minipage}
\end{center}
\begin{table}[htbp]
  \centering
  \caption{SAS Data Set d2008}

\resizebox{5in}{!}{%
  \begin{tabular}{rrrrrrrrr}
    \addlinespace
    \toprule
    home  & Game Date & away  & home\_score & away\_score & fcs   & H     & A     & home\_win \\
    \midrule
    Ball St. & 8/28/2008 & Northeastern & 48    & 14    & 1     & 1     & -1    & 1 \\
    Baylor & 8/28/2008 & Wake Forest & 13    & 41    & 0     & 1     & -1    & 0 \\
    Buffalo & 8/28/2008 & UTEP  & 42    & 17    & 0     & 1     & -1    & 1 \\
		\vdots&\vdots&\vdots &  \vdots   & \vdots    &    \vdots  & \vdots    & \vdots   &\vdots  \\
    \bottomrule
    \end{tabular}% 
    }
  \label{tab:data}%
\end{table}%

\normalsize

\bibliographystyle{asabst}
\bibliography{jqas}

%\appendix

\end{document}